%% file: nips2023.tex
\documentclass{article}

 \usepackage[nonatbib,final]{neurips_2023}

\usepackage[utf8]{inputenc} 
\usepackage[T1]{fontenc}    
\usepackage{hyperref}       
\usepackage{url}            
\usepackage{booktabs}       
\usepackage{amsfonts}       
\usepackage{nicefrac}       
\usepackage{microtype}      
\usepackage{xcolor}         
\usepackage{amsmath}
\usepackage{paralist}

\usepackage{amsfonts}
\usepackage{multirow}
\usepackage{tabularx}
\usepackage{booktabs}
\usepackage{subfig}
\usepackage{graphicx}
\usepackage{algorithm}
\usepackage{algorithmic}
\usepackage{arydshln}

\title{DeWave: Discrete EEG Waves Encoding for Brain Dynamics to Text Translation}

\author{
Yiqun Duan$^{1}$\thanks{is the first author,  $\dagger$ is the corresponding author}, \quad Jinzhao Zhou$^{1}$, \quad Zhen Wang$^{2}$, \quad Yu-Kai Wang$^{1}$, \quad Chin-Teng Lin$^{1 \dagger}$\\
  $^{1}$GrapheneX-UTS HAI Centre, Australian Artificial Intelligence Institute,\\
  Faculty of Engineering and Information Technology\\
  University of Technology Sydney, Ultimo, NSW 2007 \\
  $^{2}$School of Computer Science, The University of Sydney, Camperdown NSW 2050\\
  \texttt{ \{yiqun.duan, jinzhao.zhou\}@student.uts.edu.au,zwan4121@uni.sydney.edu.au } \\
  \texttt{ yukai.wang@uts.edu.au, chin-teng.Lin@uts.edu.au} \\
}

\begin{document}

\maketitle

\begin{abstract}
The translation of brain dynamics into natural language is pivotal for brain-computer interfaces (BCIs).
With the swift advancement of large language models, such as ChatGPT, the need to bridge the gap between the brain and languages becomes increasingly pressing. Current methods, however, require eye-tracking fixations or event markers to segment brain dynamics into word-level features, which can restrict the practical application of these systems. 
To tackle these issues, we introduce a novel framework, DeWave, that integrates discrete encoding sequences into open-vocabulary EEG-to-text translation tasks. DeWave uses a quantized variational encoder to derive discrete codex encoding and align it with pre-trained language models. This discrete codex representation brings forth two advantages: 1) it realizes translation on raw waves without marker by introducing text-EEG contrastive alignment training, and 2) it alleviates the interference caused by individual differences in EEG waves through an invariant discrete codex with or without markers.
Our model surpasses the previous baseline (40.1 and 31.7) by $3.06\%$ and $6.34\%$, respectively, achieving 41.35 BLEU-1 and 33.71 Rouge-F on the ZuCo Dataset. This work is the first to facilitate the translation of entire EEG signal periods without word-level order markers (e.g., eye fixations), scoring 20.5 BLEU-1 and 29.5 Rouge-1 on the ZuCo Dataset.
\end{abstract}

\begin{figure}[h]
\vspace{-5pt}
    \centering
    \begin{minipage}[]{0.11\linewidth}
    \centering
      \includegraphics[width=2.5cm]{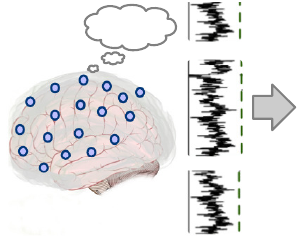}
    \end{minipage}
    \hfill
    \begin{minipage}[]{0.82\linewidth}
    \small  
    \resizebox{0.95\textwidth}{!}{
    \begin{tabular}{p{1cm}p{10cm}}
    \toprule
    \textbf{Ground Truth} & Bob attended \underline{\textbf{the University of} Texas \textbf{at}}  \underline{\textbf{Austin} where he graduated}, \underline{\textbf{Phi Beta}} \underline{\textbf{Kappa}} with a Bachelor's degree in \underline{Latin \textbf{American Studies in} 1973}, taking only two and \textbf{a half years} to \underline{complete his work}, and obtaining generally \textbf{excellent grades.}\\ \midrule
    \textbf{Predict} & was \underline{\textbf{the University of} California \textbf{at Austin}} in \underline{where he studied in \textbf{Beta Kappa}} in a degree of degree in \underline{history \textbf{American Studies in} 1975}. and a one classes \textbf{a half years} to \underline{complete the degree}. and was a \textbf{excellent grades.} \\ \bottomrule
    \end{tabular}
    }
    \end{minipage}
    \vspace{-3pt}
    \caption{Overall illustration of translating EEG waves into text~\protect\footnotemark through quantised encoding.  \label{fig:coverset}} 
    \vspace{-10pt}
\end{figure}
\footnotetext{\textcolor{black}{This visualization still keeps teacher-forcing evaluation for a fair comparison with previous methods.}}

\section{Introduction}
Decoding brain states into comprehensible representations has long been a focal point of research~\cite{kobler2022spd,NEURIPS2019_d464b5ac,duan2023cross,zhou2023belt}. Electroencephalogram (EEG) signals are particularly favored by researchers due to their non-invasive nature and ease of recording~\cite{pan2022matt,wagh2022evaluating}. Traditional EEG decoding techniques largely focus on classifying brain states into restricted categories like Motor Imaginary (MI)~\cite{saa2013discriminative,duan2022cross}, Emotion~\cite{jenke2014feature,wang2021review}, Robotic Control~\cite{BCI_wheelchair,zhou2022generalizing}, and Gaming~\cite{BCI_gaming,liao2012gaming}. However, these labels, bound to specific tasks, are insufficient for broad-based brain-computer communication. Consequently, there has been a surge of interest in brain-to-text (speech) translation in recent years.
As the current trend leans towards large language models (LLM)~\cite{bubeck2023sparks,floridi2020gpt,ouyang2022training} showcasing increasingly generalized intelligence capabilities, it becomes crucial to delve into ways of bridging the gap between brain signals and natural language representation. However, this area remains under-explored.

The early work in brain-to-text translation~\cite{herff2015brain,anumanchipalli2019speech,makin2020machine,sun2019towards} relied on external event markers like handwriting or eye-tracking fixations to segment whole brain signals into fragmentary features. This methodology treated the task as word-level classification on a small, closed vocabulary set, with each time step analyzed individually. Both invasive~\cite{herff2015brain,sun2020brain2char} (ECoG) and non-invasive~\cite{panachakel2021decoding} (EEG) brain signals have been used in these approaches. Notably, utilizing handwriting as event markers on invasive signals has led researchers~\cite{willett2021high} to achieve state-of-the-art (SOTA) recognition accuracy on closed-set character-level recognition. Wang~\cite{wang2022open} expanded the vocabulary size substantially and demonstrated the feasibility of open-vocabulary brain-to-text translation by employing pre-trained language models with word-level EEG features. However, limitations persist. The order of event markers, particularly eye-tracking fixations used to segment EEG waves into word-level features, may not match the natural word order in language~\footnote{For example, the \href{https://osf.io/2urht/}{ZuCo} Dataset collects data by simultaneously recording eye-tracking fixation and brain waves during reading tasks. However, the order of eye-fixations and spoken words may not always coincide.}. Moreover, current methods do not have the capacity for direct text translation.

In this paper, we present Discrete EEG Waves Encoding for Brain Dynamics to Text Translation (DeWave), a pioneering framework depicted in Figure~\ref{fig:dewave_model}. DeWave uses a vector quantized variational encoder to transform EEG waves into a discrete codex, linking EEG waves to tokens based on their proximity to codex book entries. This method offers two key advantages: it addresses significant distribution variances in EEG waves across individuals~\cite{onton2006imaging,saha2020intra,duan2023domain}, and rectifies order mismatches between raw wave sequences and text without eye-tracking markers. Our raw waves encoder is guided by both self-reconstruction and a contrastive supervision alignment between text embeddings and vectorized raw waves. To navigate the challenges of training with limited parallel data, DeWave leverages large-scale pre-trained language models~\cite{devlin2018bert,radford2019language,brown2020language}, specifically employing BART~\cite{lewis2019bart}, which combines BERT's bidirectional context with GPT's left-to-right decoder. Notably, our discrete codex aligns more closely with actual language tokens than continuous EEG features, serving as an interpretable bridge between EEG input and the language model.

Experiments employ non-invasive EEG signals and data from the \href{https://osf.io/2urht/}{ZuCo} dataset~\cite{DBLP:conf/lrec/HollensteinTZL20}, a large-scale public resource that records eye-tracking and EEG during natural reading tasks.
Notably, DeWave can be generalized for both word-level EEG features and raw EEG wave translation. With a robust codex representation, this work pioneers in translating the entire time period of EEG signals without the need for word-level order markers such as eye fixations.
We assess the performance using standard translation metrics~\cite{papineni2002bleu,lin-2004-rouge}. With word-level EEG features, DeWave attains 42.8 BLEU-1 and 34.9 Rouge-1, surpassing the previous baseline (40.1 and 31.7) by $6.73\%$ and $10.09\%$ respectively on the ZuCo Dataset. For raw EEG waves without event markers, DeWave achieves 20.5 BLEU-1 and 29.5 Rouge-1.
This work's contributions can be summarized in three main points.
\begin{itemize}
    \item This paper introduces discrete codex encoding to EEG waves and proposes a new framework, DeWave, for open vocabulary EEG-to-Text translation. 
    \item By utilizing discrete codex, DeWave is the first work to realize the raw EEG wave-to-text translation, where a self-supervised wave encoding model and contrastive learning-based EEG-to-text alignment are introduced to improve the coding ability. 
    \item Experimental results suggest the DeWave reaches SOTA performance on EEG translation, where it achieves 41.35 BLUE-1 and 33.71 Rouge-1, which outperforms the previous baselines by $3.06\%$ and $6.34\%$ respectively. 
    
\end{itemize}

\input{text/related_works.tex}

\input{text/methodology.tex}

\input{text/exp.tex}

\vspace{-5pt}
\section{Limitations} 
\vspace{-5pt}
\textcolor{black}{
Despite DeWave's enhancements in EEG-to-Text translation using a discrete codex and raw wave encoding, its accuracy remains far from real-life scenarios compared to traditional language-to-language translations. 
Also, to keep a fair comparison with EEG-to-Text, this paper uses a teacher-forcing setting in evaluation. This setting eliminates accumulation error and turns the sequence decoding task into a word-level classification task given the ground truth token from the previous step. This setting is relatively easier yet we think it is still valuable as it could suggest feature extraction quality while keeping a fair comparison. }

\textcolor{black}{
Additionally, the experiments in this paper are restricted to public neural reading data, not fully aligning with the "silent speech" concept of direct thought translation from human brains. Instead, the current ZuCo dataset is collected by giving people reading stimuli. 
This paper focuses on introducing Wav2Vec formation of feature extraction on raw EEG waves and introduces discrete codex as learnable representations for the EEG-to-Text translation domain. 
One scientific problem in this domain is a better way of doing the ``silent speech" task, which we are exploring as on-going research.  
}

\vspace{-5pt}
\section{Conclusion}
\vspace{-10pt}
This paper presents DeWave, a framework for the recently proposed open-vocabulary EEG-to-Text translation task~\cite{wang2022open}, introducing the concept of discrete codex encoding. This approach brings enhancement in corpus text relevancy metrics, such as BLEU and ROUGE. DeWave also expands the task to decode raw EEG waves without the assistance of eye fixation markers. 
\textcolor{black}{
Despite these advancements, the quality of brain decoding results remains substantially low and remains teacher forcing setting for fair comparison.
The translation of thoughts directly from the brain is a valuable yet challenging endeavor that warrants significant continued efforts. 
In our ongoing work, we are exploring more reasonable settings that remove teacher forcing for both training and testing. We will also include the ``neural-feedback" mechanism in this EEG-to-Text research to enhance the scientific value of this domain. }

\section*{Acknowledgement}
This work was supported in part by the Australian Research Council (ARC) under discovery grants DP210101093 and DP220100803, and the GrapheneX-UTS Human-Centric AI Centre sponsored by GrapheneX (2023-2031).


\medskip
{
\small
\bibliographystyle{plain}
\bibliography{nips2023}
}

\appendix
\input{text/appendix}

\end{document}

%% file: text/related_works.tex
\section{Related Works}
\label{sec:relatedworks}



The key to decoding natural language from EEG signals is good representations. Existing work for EEG-to-text representation can be categorized into hand-crafted representations and deep-learning representations. 
Earlier works make use of traditional methods to obtain hand-crafted representations from a sequence of EEG signals. A continuous feature representation is extracted using methods such as statistical features \cite{zhao2015classifying}, correlation coefficient matrix from fast Fourier transform (FFT) components \cite{tanaka2005electroencephalogram,xi2022differentiable,xi2022single}, wavelet transform features \cite{jahangiri2019relative,panachakel2020novel,zhou2023belt}, and Mel-frequency cepstral coefficients (MFCCs)~\cite{cooney2018mel}. These hand-crafted representations can be used to establish a mapping between EEG segments and words using distance-based methods such as linear discriminative analysis (LDA) or Support Vector Machine (SVM).

On the other hand, deep learning methods, such as \cite{zhang2018converting}, utilize a convolutional neural network (CNN) and a Long Short-Term Memory (LSTM) to learn deep features and perform classification of user’s instructions. However, the direct representation is often insufficient to discriminate a larger number of imagined text categories. In \cite{saha2019speak}, EEG signals are first classified into six phonological categories as the intermediate state using a deep convolutional autoencoder. Then latent features are used as input for another encoder to predict a total of 11 speech tokens. 
Recently, \cite{wang2022open} proposed a framework that uses a multi-layer transformer encoder to project word-level EEG feature sequences into EEG embeddings. Then a pre-trained BART model is used to decode these embeddings into words. 

%% file: text/methodology.tex
\section{Method}

\begin{figure*} 
\centering    
\label{fig:dewave_model_sub}     
\includegraphics[width=13cm]{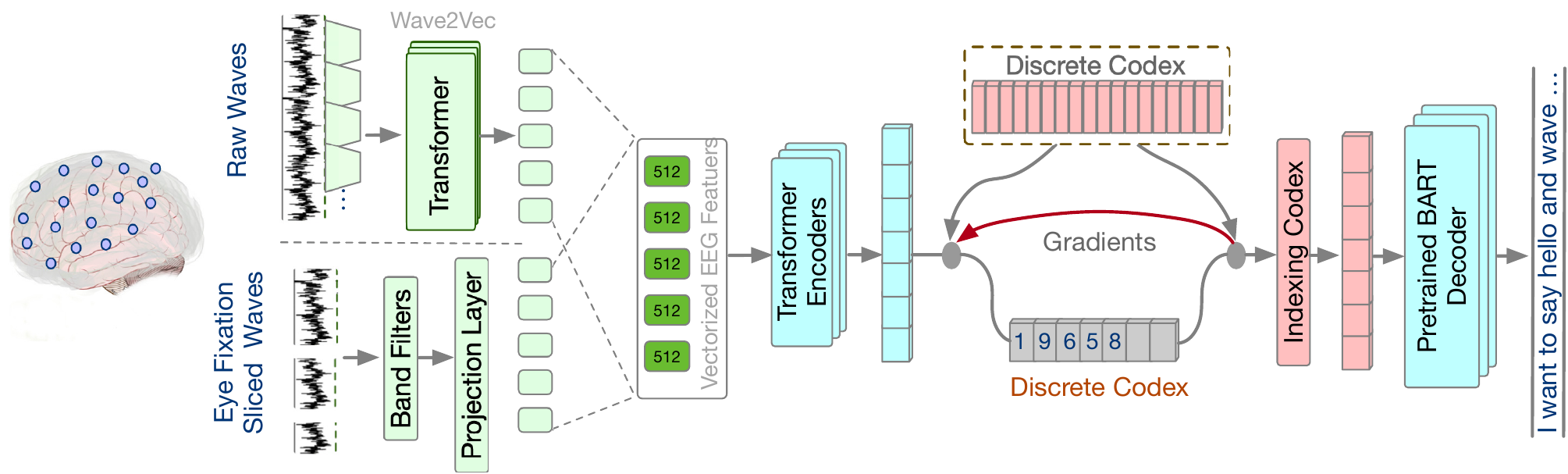}   
\vspace{-4pt}
\caption{The DeWave model structure involves vectorizing either word-level EEG features or raw EEG waves into embeddings (Section~\ref{subsec:vectorization}). The vectorized features are then encoded into a latent variable $\mathbf{z}_c(\mathcal{X})$, which is converted into a discrete latent $\mathbf{z}_q(\mathcal{X})$ through codex indexing. Finally, a pre-trained BART model translates this discrete codex representation into texts.}     
\label{fig:dewave_model}     
\end{figure*}

The overall process of DeWave is illustrated in Figure~\ref{fig:dewave_model}, where the word-level or raw EEG features are vectorized into sequences embedding and are fed into the discrete codex. The language model generates translation output based on discrete codex representation.

\subsection{Task Definition}

\label{sec:task_formulation}

Given a sequence of word-level EEG features $\mathcal{E}$, the aim is to decode the corresponding open-vocabulary text tokens $\mathcal{W}$. These EEG-Text pairs $\langle \mathcal{E}, \mathcal{W} \rangle$ are collected during natural reading, as defined in Section~\ref{sec:data}. We delve into two task settings: (1) Word-level EEG-to-Text Translation, where EEG feature sequences $\mathcal{E}$ are fragmented and re-ranked based on eye-fixation $\mathcal{F}$ aligned with each word token $\mathbf{w}$ in sequence $\mathcal{W}$; and (2) Raw EEG Waves to Text Translation, where EEG feature sequences $\mathcal{E}$ are directly vectorized into embedding sequences for translation without any event markers, a more challenging but practical real-time setting. DeWave is the pioneering work in this latter task.

\subsection{Discrete Codex}

Discrete representation is first proposed in VQ-VAE~\cite{van2017neural}. DeWave is the first work to introduce discrete encoding into EEG signal representation. 
The discrete representation could benefit both the word-level EEG features and the raw EEG wave translation. 
Introducing discrete encoding into brain waves could bring two aspects of advantages. 
1) It is widely accepted that EEG features have a strong data distribution variance across different human subjects. 
Meanwhile, the datasets can only have samples from a few human subjects due to the expense of data collection. 
This severely weakened the generalized ability of EEG-based deep learning models
By introducing discrete encoding, we could alleviate the input variance to a large degree as the encoding is based on checking the nearest neighbor in the codex book. 
2) The codex contains fewer time-wise properties which could alleviate the order mismatch between event markers (eye fixations) and language outputs. 
Meanwhile, because of this property of codex represents, DeWave is the first work that could realize the direct translation from raw EEG waves without any event marker to give the order. 

\textbf{Inference:}\quad
Given the EEG waves $\mathcal{E}$, it is firsted vectorized into embedding as introduced in Section~\ref{subsec:vectorization} with ($\mathcal{X}=\Theta(\mathcal{E}, \mathcal{F})$) or without ($\mathcal{X}=\Theta(\mathcal{E})$) eye fixations $\mathcal{F}$, where $\mathcal{X}$ is the embedding sequence. 
A codex book $\{ \mathbf{c}_i \}\in \mathbb{R}^{k \times m}$ is initialized with number $k$ of latent embedding with size $m$. 
The vectorized feature $\mathcal{X}$ is encoded into $\mathbf{z}_c(\mathcal{X})$ through a transformer encoder. 
The discrete representation is acquired by calculating the nearest embedding in the codex of input embedding $\mathbf{x} \in \mathcal{X}$ as shown in Equation~\ref{eq:inf}.
\begin{equation}\label{eq:inf}
    \begin{aligned}
    \mathbf{z}_q(\mathcal{X}) =  \{ \mathbf{z}_q(\mathbf{x}) \}_i, \quad
    \mathbf{z}_q(\mathbf{x})=  c_k,\quad k= \rm{argmin}_j \Vert {z}_c(\mathbf{x}) - \mathbf{c}_j\Vert_2
    \end{aligned}
\end{equation}
Different from the original VQ-VAE which decodes the original input, Dewave directly decodes the translation output given the representation $\mathbf{z}_q(\mathcal{X})$. 
Given a pre-trained language model, the decoder predicts text output with $P(\mathcal{W}|\mathbf{z}_q(\mathcal{X})$.

\textbf{Learn:}\quad
The codex is like a bridge connecting the vectorized EEG feature and the language model.
Compared to learning direct EEG-to-text relation, DeWave learns a better discrete codex for the language model. 
It is easier to learn since 
We learn the discrete codex by the combination of the loss functions in three parts,
\begin{equation}\label{eq:codex}
\begin{aligned}
    L_{\rm{}} =  -\rm{log}(p(\mathcal{W}|\mathbf{z}_q(\mathcal{X}))  
     + \Vert \mathbf{sg}[{\mathbf{z}}_c(\mathbf{x})] - \mathbf{z}_q(\mathbf{x})\Vert^2_2  + \beta \Vert {\mathbf{z}}_\mathbf{c}(\mathbf{x}) - \mathbf{sg}[\mathbf{z}_q(\mathbf{x})]\Vert^2_2
\end{aligned}
\end{equation}
where the loss maximize the log-likelihood of language outputs $\rm{log}(P(\mathcal{W}|\mathbf{z}_q(\mathcal{X}))$ and minimize the distance between latent variable $\mathbf{z}$ and the codex value $\mathbf{z}$. 
Here, the $\mathbf{sg}$ denotes the stop gradients. The learning is robust for $\beta$ from 0.1-2.0, where we set it as 0.2 throughout the training process.

\subsection{EEG Vectorization}
\label{subsec:vectorization}

\textbf{Word-Level EEG Features With Event Markers:} \quad
The EEG waves are first sliced into fragments according to the eye-tracking fixation of word sequences given in the annotation. 
Similar to \cite{wang2022open}, we calculate the statistical result of four frequency band filters, Theta band (5-7Hz), the Alpha band (8-13Hz), the Beta band (12-30Hz), and Gamma band (30Hz-)~\cite{mcfarland2006bci} to get the statistic frequency features of each fragment. 
It is noted that although different fragments may have different EEG window sizes, the statistical results are the same (embedding size 840). 
A multi-head transformer layer is applied to project the embedding into feature sequences with latent size 512.

\begin{figure}[t]
    \centering
    \begin{minipage}[]{0.75\linewidth}
    \includegraphics[width=10.5cm]{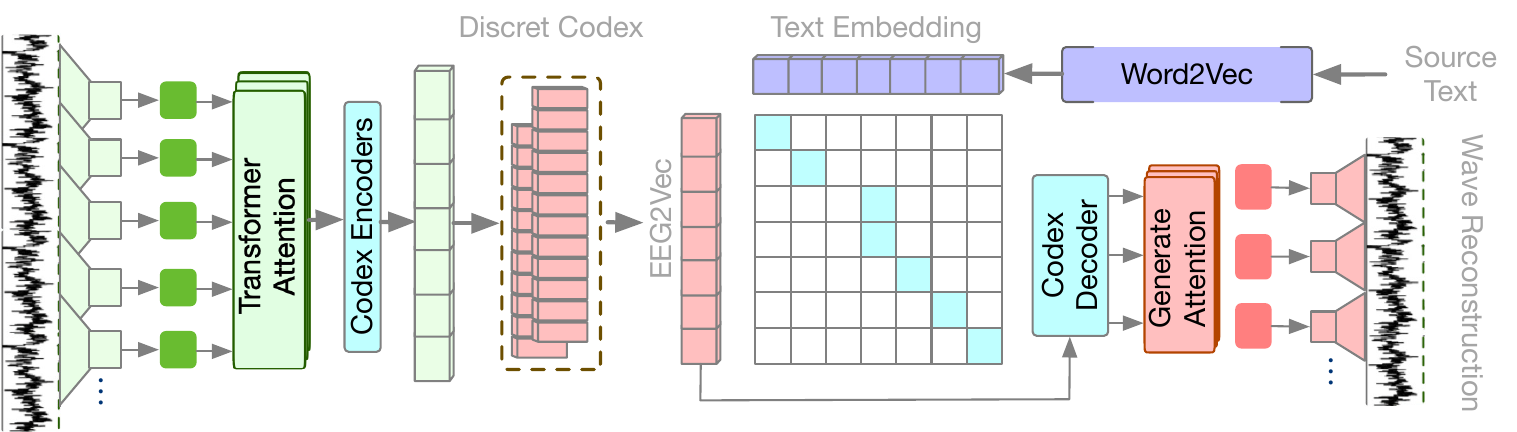}
    \end{minipage}
    \begin{minipage}[]{0.23\linewidth}
    \small
    \resizebox{1\textwidth}{!}{
    \begin{tabular}{l}
    \hline
    \multicolumn{1}{c}{\textbf{Encoder}} \\ \hline
    \multicolumn{1}{l}{\textit{ConFormer~\cite{gulati2020conformer}}} \\ 
    kernel  (10,3,3,3,2)      
    stride  (3,2,2,2,2)         \\\hline
    \textit{Codex Transformer}         \\
    heads=8, dim=512,  
    layer=6 \\ \hline
    \multicolumn{1}{c}{\textbf{Decoder}} \\ \hline 
    \textit{Codex Transformer} \\ 
    heads=8, dim=512,
    layer=6 \\ \hline
    \textit{CNNs} \\
    kernel  (3,3,3),         
    stride  (2,2,3)          \\\hline
    \textit{Transpose CNN} \\
    kernel  (3)     stride  (2)          \\ \hline
    \end{tabular}
    }
    \end{minipage}
    \vspace{-2pt}
    \caption{The image demonstrates the process of self-supervised pre-training for raw waves. The left subgraph details our strategy for directing the encoder, utilizing both self-reconstruction and text alignment through contrastive learning. 
    \label{fig:sswave2vec_clip}}
    \vspace{-6pt}
\end{figure}

\textbf{Raw EEG Waves: Self-Guided Waves to Discrete Codex}\quad
Our self-supervised EEG wave encoder transforms raw EEG signals into a sequence of embeddings~\cite{baevski2020wav2vec,zhou2023speech2eeg} as illustrated in Figure~\ref{fig:sswave2vec_clip}. It has two guiding principles: Self-Reconstruction, where the encoder is trained to transform and subsequently reconstruct the original waveforms from discrete codices; and Text Alignment, where the codices' encoding is semantically aligned with word vectors, fostering the development of text-aligned EEG signal representations.

For structure-wise, a conformer-based multi-layer encoder with specially designed hyperparameters is employed. The one-dimensional convolution layer processes the EEG waves to generate the embedding sequence~\footnote{perception field is roughly 200ms with overlap 100ms for each embedding}, fusing the EEG channels into a unique embedding for each period. We apply bi-directional transformer attention layers to the sequence to capture temporal relations.

For the reconstruction process, a decoder transformer and transpose convolution structure then convert these discrete embeddings back into raw waves.
Given the reconstruction process as $\Tilde{\mathcal{X}} = \mathcal{\phi}(\mathbf{z}_q(\mathcal{X})$, the self-supervised loss could be modified into:
\begin{equation}\label{eq:sswave}
\begin{aligned}
    L_{\rm{wave}} = \frac{1}{n}\sum( \mathcal{\phi}(\mathbf{z}_q(\mathcal{X})) -\mathcal{X})^2_n  + \Vert \mathbf{sg}[{\mathbf{z}}_c(\mathbf{x})] - \mathbf{z}_q(\mathbf{x})\Vert^2_2  + \beta \Vert {\mathbf{z}}_\mathbf{c}(\mathbf{x}) - \mathbf{sg}[\mathbf{z}_q(\mathbf{x})]\Vert^2_2,
\end{aligned}
\end{equation}
where the model calculates the mean square loss between the reconstructed wave and the wave ground truth to perform self-supervised training. 

To obtain a semantically coherent codex, we introduce a cross-modality contrastive learning approach distinct from prevalent methods.
Unlike CLIP~\cite{radford2021learning,li2022exploring,li2023clip} that contrasts CLS embeddings between sample pairs within a mini-batch, our approach operates within a single EEG-text pair.
We contrast the EEG-tokenized codex embeddings sequence $z_q$ with the text embeddings sequence $z_t$. 
Assuming the raw wave feature extractors can produce a token sequence in an \textbf{organized chronological order}, we treat the diagonal EEG codex and text word2vec encoding pairs as positive pairs within the sequences. All other pairs are considered negative.
The model is trained to minimize the distance between embeddings of positive pairs and maximize that of the negative pairs.
For a given EEG-text pair $(i, j)$, the loss is defined in Equation~\ref{eq:contrast}
\begin{equation}\label{eq:contrast}
    L_{\rm{contrast}} = -\frac{1}{n}\sum log\left[\frac{\exp(\mathbf{s}_{ii} / \tau)}{\sum_{k=1}^{N} \exp(\mathbf{s}_{ik} / \tau)}\right], \mathbf{s}_{ij} = \mathbf{z}_q(\mathbf{x}^i)^T \mathbf{z}_t(j)
\end{equation}
Here, $\tau$ is the temperature parameter, $N$ is the total number of EEG-text pairs in a batch, and the sum in the denominator is over all $N$ EEG-text pairs and $N$ text-EEG pairs. 
The EEG embeddings are expected to correctly match with their corresponding text while distinguishing them from mismatched EEG-text pairs.
The total loss then becomes a combination of the original Wave2Vec loss and the contrastive loss as $L_{\rm{total}} = L_{\rm{wave}} + \alpha L_{\rm{contrast}}$.

By this means, the model not only learns to reconstruct the EEG signal but also learns a robust representation of the signal that aligns with the corresponding text embeddings. This cross-modal learning can potentially improve the translation system by bridging the gap between EEG signals and the semantic content of the text.

\subsection{Language Model}

We used large-scale text corpus pre-trained BART~\cite{lewis2019bart} as the generative language model for translation output. 
As the EEG-to-text translation data is quite limited, leveraging the BART model could introduce prior knowledge of text relations. 
In that case, the translation system only needs to learn a codex representation for the language model, which is easier to learn. 
The codex representations are fed into pre-trained~\footnote{\href{https://huggingface.co/facebook/bart-large}{https://huggingface.co/facebook/bart-large}} BART model and get the output hidden states. 
A fully connected layer is applied on the hidden states to generate English tokens from pre-trained BART vocabulary $\mathcal{V}$.

\subsection{Training Paradigm}
DeWave is trained through a multi-stage process, where the training process is illustrated in Appendix~\ref{ap:training}. 
In the first stage, we do not involve the language model in weight updates. 
The target of the first stage is to train a proper encoder projection $ \theta_{codx}$ and a discrete codex representation $\mathcal{C}$ for the language model. 
In the second stage, the gradient of all weights, including language model $\theta_{BART}$ is opened to fine-tune the whole system.

%% file: text/exp.tex
\section{Experiments}

\subsection{Dataset}\label{sec:data}
DeWave utilize both ZuCo 1.0~\cite{hollenstein2018zuco} and 2.0~\cite{hollenstein2020zuco} for experiments. The dataset simultaneously recorded the text and EEG corpus during Normal Reading (NR) and Task-Specific Reading (TSR) tasks. 
The EEG waves are collected with a 128-channel system under a sampling rate 500Hz through a frequency band filter from 0.1Hz to 100Hz. 
However, after the noise canceling process, only 105 channels~\cite{hollenstein2018zuco} are used for translation. 
Similar to ~\cite{wang2022open}, we slice the EEG wave according to the eye fixation and calculate the frequency features.  
For raw EEG waves, the signal is normalized into a value range of $0$-$1$ for decoding. 
The reading task’s data are divided into the train ($80\%$), development ($10\%$), and test ($10\%$) respectively by 10874, 1387, and 1387 unique sentences with no intersections. 
Please refer to Appendix~\ref{ap:dataset} for a detailed description.

\subsection{Implementation Details }
For word-level EEG features, we use the $56$ tokens each with an $840$ embedding size. For raw EEG waves, we clip or pad the EEG waves up to sample point 5500 with a constant value of zero. A transformer layer with head number 8 and a $1\times1$ convolutional layer are combined to fuse multiple EEG channels into an embedding sequence with size 512. 
DeWave uses a codex with size 2048 where each codex latent is an embedding with size 512. The ablation study (Section~\ref{subsec:abl}) gives a discussion about the codex size. 
All models are trained on Nvidia V100 and A100 GPUs. 
For the self-supervised decoding for raw waves, we use a learning rate of 5e-4 and a VQ coefficient of 0.25 for training 35 epochs. 
For training the codex (stage 1), DeWave uses a learning rate of 5e-4 for 35 epochs. 
For finetuning the translation (stage 2), DeWave uses a learning rate of 5e-6 for 30 epochs. 
We use the SGD as the optimizer for training all the models. 
Due to limited space, refer to Appendix~\ref{ap:implementation} for more details.

\subsection{Evaluation Metrics}
\label{subsec:metrics}

We evaluate translation performance using NLP metrics, BLEU and ROUGE, as shown in Table~\ref{tb:score}. For word-level EEG features, we compare our results to EEG-to-Text~\cite{wang2022open}, maintaining a consistent language model for fairness. In the absence of methods for raw EEG waves, we establish a baseline EEG-to-Text$^{\dagger}$ by segmenting the entire EEG waves into a sequence embedding using a 200ms time window with a 100ms overlap. We adapt Wave2Vec, originally developed for speech recognition, to brain waves and compare it with our approach, DeWave. Furthermore, we adapt unsupervised raw EEG waves classification methods BENDR~\cite{kostas2021bendr} and SCL~\cite{jiang2021self}, using SSL pre-training and feature extraction for comparison, underscoring the impact of discrete encoding.

\input{text/table_score.tex}

\textbf{Word-Level EEG Features: \quad}
For the word-level EEG feature, we observe that introducing the discrete brain representation could help DeWave reach BLEU-\{$1,2,3,4$\} scores of $41.35, 24.15, 13.92$, and $8.22$, which respectively outperform the previous baseline by $1.23~(+3.06\%)$, $0.97~(+4.18\%)$, $1.31~(+10.38\%)$ and $1.54~(+18.73\%)$. 
It is observed that the increasing ratio is more significant for larger grams evaluation. 
DeWave achieve ROUGE-1 score $28.84$ (R), $31.69$ (P), and $30.10$ (F) which outperforms the previous baseline by $-0.02~(-0.06\%)$,  $1.98~(+6.35\%)$, and $0.59~(+1.9\%)$. 
The increase of the matrix is higher for longer phrases (3-gram and 4-gram). 
This is rational since the single-word level EEG features are sliced by eye fixation during reading, which may naturally contain noises since the human subject may not really think about the corresponding words when looking at them. 

\textbf{Raw Waves: \quad}
DeWave represents an unprecedented endeavor in translating raw EEG waves directly, obviating the need for event markers. This is achieved through the application of a learned Wave2Vec model, which converts waves into discrete codex tokens.
When benchmarked against the baseline method - which merely slices the wave according to a time window - DeWave exhibits marked improvement. It attains BLEU-{$1,2,3,4$} scores of $20.51, 10.18, 5.16$, and $2.56$, respectively, surpassing the baseline by margins of $7.44~(+56.92\%)$, $4.44~(+76.1\%)$, $2.61~(+102.35\%)$, and $1.42~(+129.09\%)$. This underscores the superiority of a learned discrete embedding representation over the rudimentary time-window baseline.
In comparison with cutting-edge self-supervised learning (SSL) methods, DeWave consistently outperforms them, including contrastive learning (CL)-based methods such as SCL~\cite{jiang2021self}, the original Wave2Vec~\cite{baevski2020wav2vec}, and BENDR~\cite{kostas2021bendr}.

\subsection{Cross-Subject Performance}
\begin{figure}[hbpt]
\hspace{-0.6cm} 
\begin{minipage}[]{0.245\linewidth}
\includegraphics[width=4.1cm]{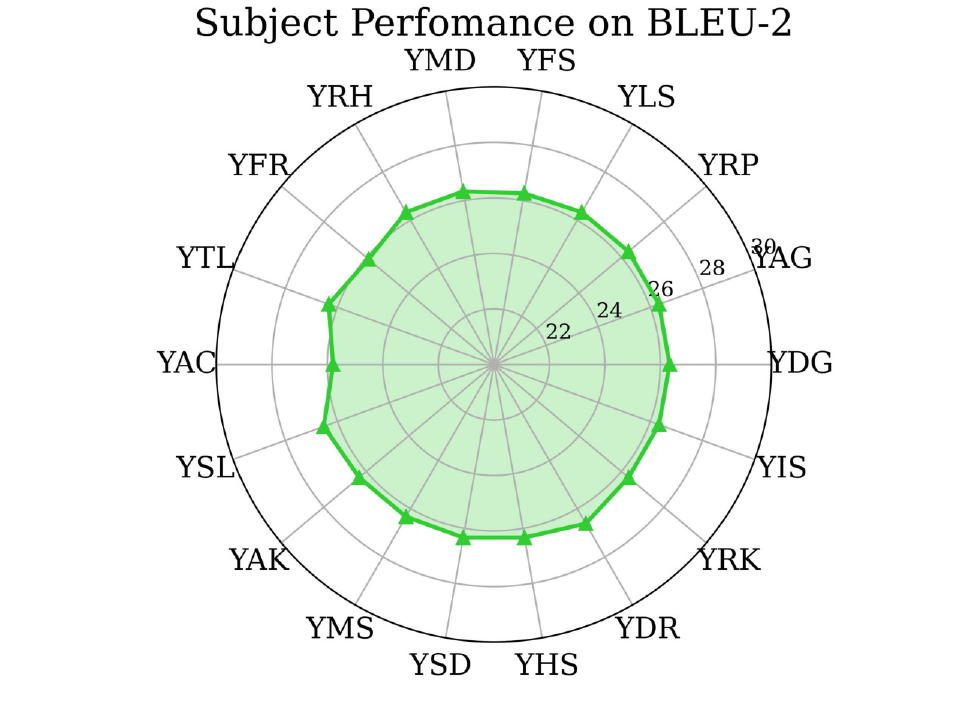}
\end{minipage}
\begin{minipage}[]{0.245\linewidth}
\includegraphics[width=4.1cm]{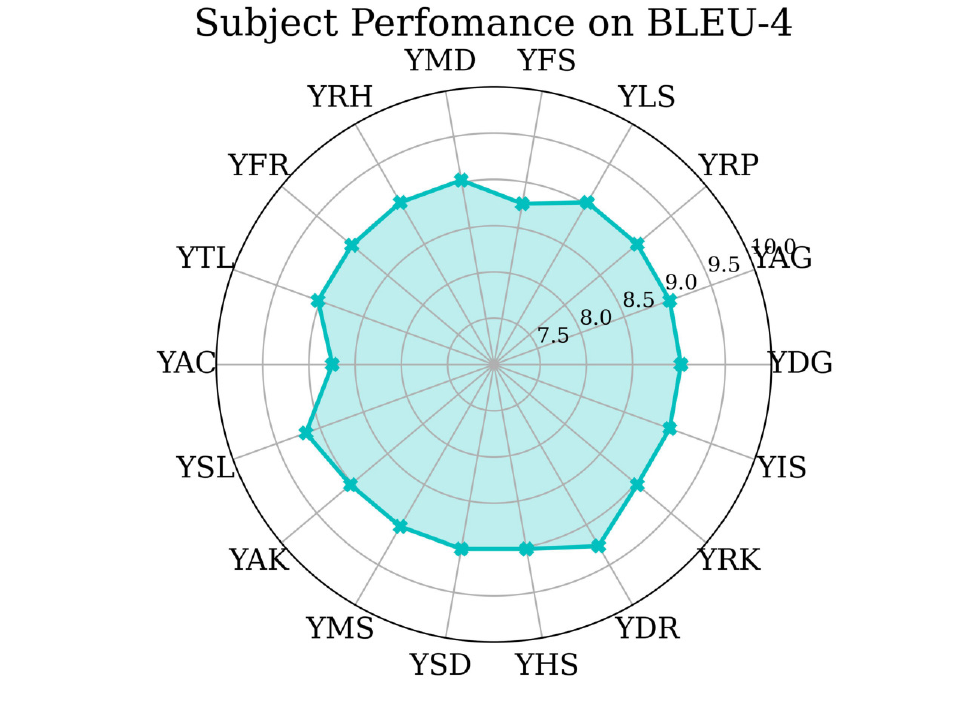}
\end{minipage}
\begin{minipage}[]{0.245\linewidth}
\includegraphics[width=4.1cm]{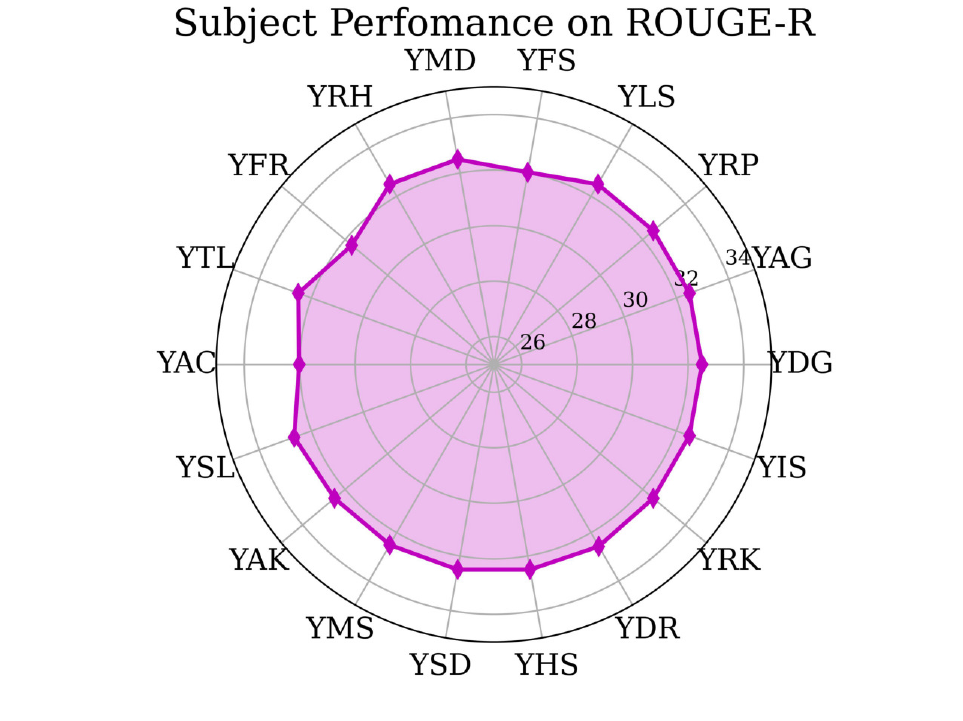}
\end{minipage}
\begin{minipage}[]{0.25\linewidth}
\includegraphics[width=4.1cm]{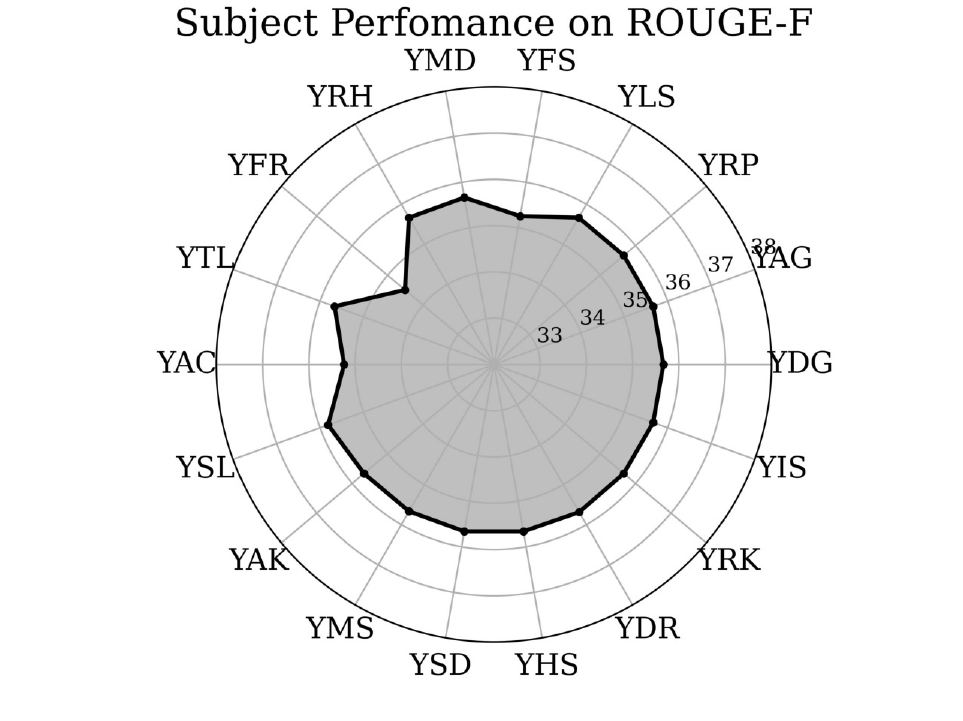}
\end{minipage}
\begin{minipage}[]{0.25\linewidth}
\end{minipage}
\vspace{-7pt}
\caption{The cross-subjects performance on ZuCo dataset. \label{fig:crossradar}}
\vspace{-10pt}
\end{figure}

Cross-subject performance is of vital importance for practical usage. To further illustrate the performance variance on different subjects, we train the model by only using the data from subject YAG and test the metrics on all other subjects. The results are illustrated in Figure~\ref{fig:crossradar}, where the radar chart denotes the performance is stable across different subjects. Please refer to Appendix~\ref{sec:subjectwise_ap} for more detailed results. 


\subsection{Generated Samples}

In Table~\ref{tab:generation_results}, we display visualized examples of text generated from unseen EEG signals. Despite the challenge of thought translation and limited prior research, our model yields meaningful results, aligning keywords and forming similar sentence structures, although it may not yet match traditional language translation tasks.

\input{text/table_targets.tex}

The model is more adept at matching verbs than nouns. For instance, \textbf{implies}'' vs. \textbf{implies}'', \textbf{the author who wrote it}'' vs. \textbf{the man who wrote it}'' both effectively convey the intended sentiment of the sentence. However, when it comes to nouns, we observe a tendency towards synonymous pairs rather than precise translations, such as \underline{the man}'' vs. \underline{the author}'', \underline{Burroughs}'' vs. \underline{Heroughs}'', \underline{edition}'' vs. \underline{version}''.
Our analysis suggests two potential causes for this. First, when the brain processes these words, semantically similar words might produce similar brain wave patterns. Given the inherent noise in brain waves, the codex might group these features under the same value. Second, the volume of EEG-to-Text pairs available for training is significantly smaller than that for traditional language translation. Hence, some degree of error in translating unseen nouns or sentences is to be expected.

Translation on raw EEG waves is naturally harder than word-level translation, as it lacks eye fixation to suggest the relationship between the period of waves and the word target. 
Table~\ref{tab:generation_results} meet our expectation that the results on raw EEG waves are not as good as those on word-level features, especially on the real semantic meaning of the sentence (sample (2) and sample (5) have the same target for comparison).
However, the translation still could output the correct translation of certain words, such as ``\textbf{much of}'' vs. ``\textbf{much of}'', ``\textbf{individual}'' vs. ``\textbf{individual}'', and ``\underline{more complicated story}'' vs. ``\underline{more exciting thing}''.
Although EEG-to-text is a hard topic, DeWave suggests the feasibility of translation improvements. 

\subsection{Ablation Study}\label{subsec:abl}

\begin{figure}[hbpt]
\vspace{-5pt}
    \begin{minipage}[]{0.3\linewidth}
    \includegraphics[width=4cm]{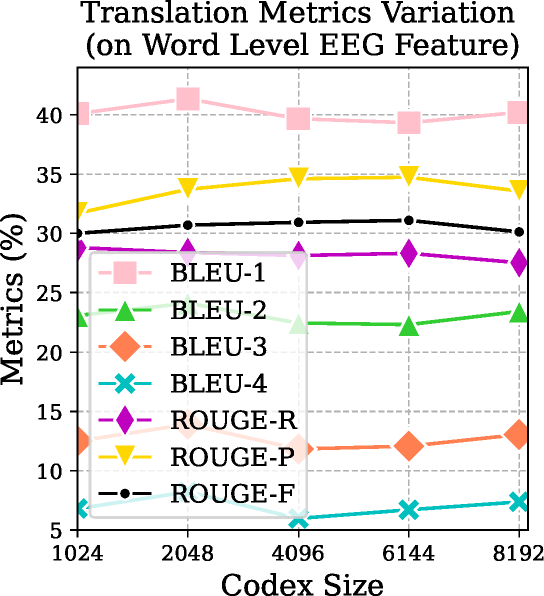}
    \end{minipage}
    \begin{minipage}[]{0.3\linewidth}
    \includegraphics[width=4cm]{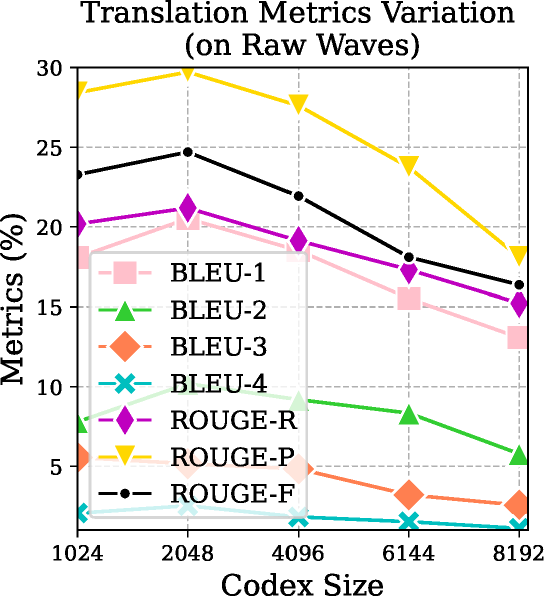}
    \end{minipage}
    \begin{minipage}[]{0.40\linewidth}
    \includegraphics[width=5cm]{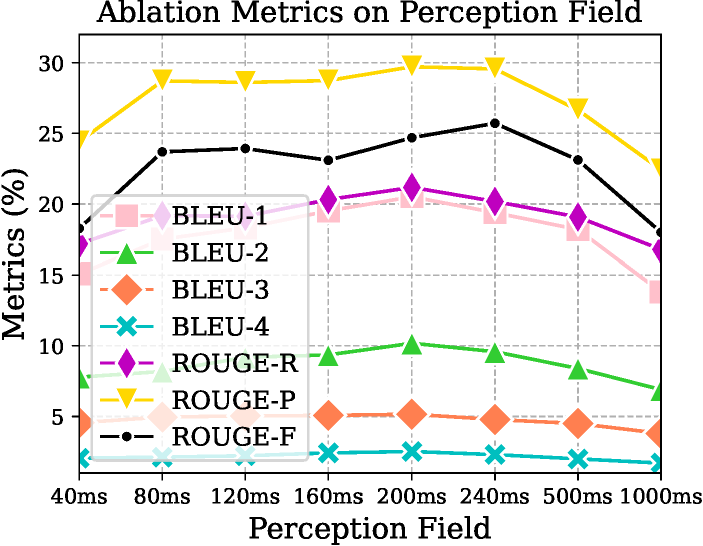}
    \end{minipage}
    \vspace{-3pt}
    \caption{Ablation study on different codex sizes and perception fields (raw waves). \label{fig:codexabl}}
    \vspace{-10pt}
\end{figure}

\textbf{Discrete Codex}\quad DeWave encodes EEG waves into a discrete codex, aiming for a language model-friendly representation. We evaluated performance against varying codex sizes (1024 to 8192) to ascertain if larger sizes yield better results. As depicted in Figure~\ref{fig:codexabl}, we found no strong correlation between codex size and model performance. A codex size of 2048 yielded the highest BLEU score average, and while the ROUGE score slightly improved with larger sizes, there was no clear evidence that increasing codex size consistently enhanced performance.

Raw EEG wave performance fluctuates noticeably with codex size. Performance improves when increasing the codex size from 1024 to 2048, but any larger size reduces performance. This variation may result from the training formation; word-level EEG data, selected by eye fixations, contains less noise than raw waves. We hypothesize that our current training data may be insufficient for larger codex sizes. Additionally, our experiments indicate that the latent dimension of the codex doesn't significantly affect performance.

\textbf{Perception Time Window}\quad
We also conduct the ablation study on the model structure for the wave2vec model illustrated in Figure~\ref{fig:codexabl}. 
As the model utilize a multi-layer CNN model to slide through the raw waves, the model compresses the waves for perception. 
The compress ratio decided how large the perception field is for each extracted embedding feature. 
As described in Section~\ref{subsec:vectorization}, the model utilized a perception field of 200ms with an overlap of 100ms. 
We conduct an ablation study of different perception fields and report it in Figure~\ref{fig:codexabl}
Where it is observed that the model performance is significantly lower when the perception field is smaller than 80ms or larger than 240ms.
The model could achieve similar results in the perception field 120ms to 240ms. 
The model reaches a small peak around 200ms to 240ms. 
We think this phenomenon is rational since the normal reading speed for humans is around 160-400 words per minute (reading speed may vary from different material and human subjects). 
In other words, the reading period for each word is 150-375ms on average, which roughly meets our observation between 200ms-240ms.


\textbf{Self-Supervision Initialization}\quad Theoretically, we could pre-train the codex by introducing a decoder and calculate the reconstruction loss with the original input for both word-level EEG and raw EEG waves. 
We prefix other parameters as our best setting with codex size 2048 and compare the impact in Table~\ref{tb:albss}.
\input{text/table_ss_abl.tex}
For word-level EEG features, the impact of pre-train is mostly on BLEU scores while the ROUGE score does not have much variance. 
Without pre-train, the BLEU-$\{1,2,3\}$ respectively drop by $0.64$, $1.21$, and $1.52$. 
For direct translation on raw waves, the impact is significantly larger. 
Without self-supervised initialization, the BLEU-$\{1,2,3\}$ respectively drop by 3.93 $(\downarrow 19.16\%)$, 2.40 $(\downarrow 23.57\%)$, and 1.48 $(\downarrow 28.68\%)$.
Similar observation also appears on ROUGE scores. 
This phenomenon is rational since raw wave decoding requires the model to pick useful features without any help from eye fixations. 
The self-supervised initialization could help the model form a preliminary ability to extract time-wise or channel-wise features from raw waves. 


%% file: text/table_score.tex
\begin{table*}[hbpt]
\centering
\caption{\label{tb:score} Evaluation metrics of EEG-to-Text translation under both word-level features input and raw waves input, where \small{+Contrastive} denotes boost encoder with $L_{contrast}$ mentioned in Sec.~\ref{subsec:vectorization}.\textcolor{black}{ For a fair comparison, these results keep the same teacher-forcing evaluation setting as EEG-to-Text~\cite{wang2022open}.} }
\vspace{-3pt}
\small
\resizebox{0.95\textwidth}{!}{
\begin{tabular}{cllllllll}
\toprule
\multirow{2}{*}{\begin{tabular}[c]{@{}l@{}}\textbf{Source}\end{tabular}} & \multirow{2}{*}{\textbf{Method}} & \multicolumn{4}{c}{\textbf{BLEU-N} ($\%$)}                                      & \multicolumn{3}{c}{\textbf{ROUGE-1} ($\%$)}                                           \\ \cline{3-9}
&                         & \multicolumn{1}{c}{N=1} & \multicolumn{1}{c}{N=2} & \multicolumn{1}{c}{N=3} & \multicolumn{1}{c}{N=4} & \multicolumn{1}{c}{R} & \multicolumn{1}{c}{P} & \multicolumn{1}{c}{F} \\ \hline
\multirow{2}{*}{\begin{tabular}[c]{@{}l@{}}\textbf{Word-level} \textbf{features}\end{tabular}} & EEG-to-Text~\cite{wang2022open}         & 40.12                   & 23.18                   & 12.61                   & 6.80                    & \textbf{28.84}        & 31.69                 & 30.10                 \\
 & DeWave                  & \textbf{41.35}          & \textbf{24.15}          & \textbf{13.92}          & \textbf{8.22}           & 28.82                 & \textbf{33.71}        & \textbf{30.69}        \\ \hline
\multirow{6}{*}{\begin{tabular}[c]{@{}l@{}}\textbf{Raw} \textbf{waves}\end{tabular}}  & EEG-to-Text$^{\dagger}$~\cite{wang2022open}       & 13.07                   & 5.78                    & 2.55                    & 1.10                    & 15.22                 & 18.08                 & 16.36                 \\
& Wave2Vec~\cite{baevski2020wav2vec}        & 18.15                   & 8.94                   & 3.89                 & 2.04                    & 18.96                & 23.86               & 20.07                \\
& BENDR~\cite{kostas2021bendr}        & 18.48                   & 9.16                    & 4.05                  & 2.15                    & 19.03                & 25.22                & 21.18                \\
& DeWave                  & \textbf{20.51}          & \textbf{10.18}          & \textbf{5.16}           & \textbf{2.52}           & \textbf{21.18}        & \textbf{29.42}        & \textbf{24.27}        \\ 
& DeWave+Contrastive & \textbf{21.09}          & \textbf{10.69}          & \textbf{5.88}           & \textbf{3.04}           & \textbf{22.01}        & \textbf{29.95}        & \textbf{24.68}        \\ 
\bottomrule
\end{tabular}
}
\end{table*}

%% file: text/table_targets.tex
\begin{table*}[hbpt]
    \small
    \centering
    \caption{Translation results on the unseen EEG waves, where \textbf{bold} denotes a correct match between ground truth and our prediction. \underline{Underline} denotes a fuzzy match with similar semantic meanings. 
    \textcolor{black}{
    For a fair comparison, these results keep the same teacher-forcing evaluation setting as EEG-to-Text~\cite{wang2022open}. This means the decoding process eliminates accumulated errors and just predicts the current token with the GT token from the last step.  } 
    \label{tab:generation_results} }
    \vspace{-4pt}
    \resizebox{0.95\textwidth}{!}{
    \begin{tabular}{l p{14cm} }
    \toprule 
    \multicolumn{2}{c}{\textbf{Decoding Results with Eye Fixation Assistance}} \\ \midrule
     \multirow{2}{*}{(1)} 
     & Ground Truth: Everything its title \textbf{implies}, a standard-\textbf{issue} crime drama spat out from the Tin\underline{seltown} assembly line...\\
     \cmidrule{2-2}
     & Prediction: is own \textbf{implies}, including great of \textbf{issue}, novel of the beginning\underline{seltowns}....\\
     \midrule
     
     \multirow{2}{*}{(2)} 
     & Ground Truth: \underline{``The Kid Stays in the Picture''} \underline{is a great story}, terrifically told by \textbf{\underline{the man} who wrote it} but this Cliff Notes \underline{edition} \textbf{is a cheat}.\\
     \cmidrule{2-2}
     & Prediction: The film \underline{"says in the Game''} \underline{is a film} about but movie was written, \textbf{\underline{the author} who wrote it}. also its \underline{version} \textbf{is a cheat}.\\
     \midrule
     
     \multirow{2}{*}{(3)} 
     & Ground Truth: During \textbf{Kerouac's time} at Columbia University, \textbf{\underline{Burroughs} and Kerouac} got into trouble with \underline{the law} for failing to report \textbf{\underline{a} murder}; this incident \textbf{\underline{formed} the basis of} a mystery novel ...\\
     \cmidrule{2-2}
     & Prediction: \textbf{Keouac's time} at the, , \textbf{\underline{Heroughs} and Kerouac} were along a for \underline{the police} for their to pay \textbf{\underline{the} murder}. they \textbf{\underline{led led} the basis of} the lawsuit ...\\
    \midrule
     
     \toprule 
     
     \multicolumn{2}{c}{\textbf{Decoding Results with Raw Waves}} \\ \midrule
     
     \multirow{2}{*}{(4)} 
     & Ground Truth: Every \textbf{individual} will see the movie through the prism of \underline{his or her own beliefs} and prejudices ...\\
     \cmidrule{2-2}
     & Prediction: Everyday \textbf{individual} is their results. their eyes of \underline{his or her own personal}. desires. and ... 
     \\
     \midrule
     
     \multirow{2}{*}{(5)} 
     & Ground Truth: \underline{``The Kid Stays in the Picture''} is a great story, terrifically told by the man who wrote it but this Cliff Notes edition \underline{is a cheat}.\\
     \cmidrule{2-2}
     & Model Output: \underline{The Price'says in the Middle}. and a common deal. and for good. this moment. made it \underline{is still little}.\\ \midrule
     
     \multirow{2}{*}{(6)} 
     & Ground Truth: \textbf{much of} this well-acted but dangerously slow thriller feels \textbf{like} a preamble to a bigger, \underline{more complicated story}, \\
     \cmidrule{2-2}
     & Model Output: \textbf{much of} this is-being but not over-. \textbf{like} it disaster-ble to a new and \underline{more exciting thing}, \\ 
     
    
    \bottomrule
    \end{tabular}
    }
    \vspace{-15pt}
\end{table*}


%% file: text/table_ss_abl.tex
\begin{table}[hbpt]
\vspace{-10pt}
\caption{Ablation on self-supervised pre-trained codex weights. \label{tb:albss}}
\small
\centering
\resizebox{0.6\textwidth}{!}{
\begin{tabular}{lp{0.5cm}p{0.5cm}p{0.5cm}p{0.6cm}p{0.6cm}p{0.6cm}p{0.6cm}}
\toprule
&\textbf{Pretrain} & \multicolumn{3}{c}{\textbf{BLEU} ($\%$)} & \multicolumn{3}{c}{\textbf{ROUGE-1} ($\%$)} \\ \cline{3-8} 
& &\multicolumn{1}{c}{N=1} & \multicolumn{1}{c}{N=2} & \multicolumn{1}{c}{N=3} & \multicolumn{1}{c}{R} & \multicolumn{1}{c}{P} & \multicolumn{1}{c}{F} \\ \midrule
\multirow{2}{*}{\begin{tabular}[c]{@{}l@{}}\textbf{Word-level} \\\textbf{features}\end{tabular}} & $\checkmark$ & 41.35 & 24.15 & 13.92 & 28.82 & 33.71 & 30.69 \\
& $\times$ & 40.71 & 22.94 & 12.40 & 28.01 & 34.08 & 30.63 \\ \midrule
\multirow{2}{*}{\begin{tabular}[c]{@{}l@{}}\textbf{Raw} \textbf{waves}\end{tabular}} & $\checkmark$ & 20.51 & 10.18 & 5.16 & 21.18 & 29.42 & 24.27 \\
& $\times$ & 16.58 & 7.78 & 3.68 & 17.86 & 19.84 & 18.33 \\ 
\bottomrule
\end{tabular}
}
\end{table}

%% file: text/appendix.tex
\newpage
\vspace{7pt}
\section*{\Large \centering Supplementary Material for DeWave: \\ 
\vspace{7pt}
{Discrete Encoding of EEG Waves for EEG to Text Translation}}

\vspace{20pt}

\begin{figure}[h]
    \centering
    \begin{minipage}[]{0.95\linewidth}
    \centering
      \includegraphics[width=7cm]{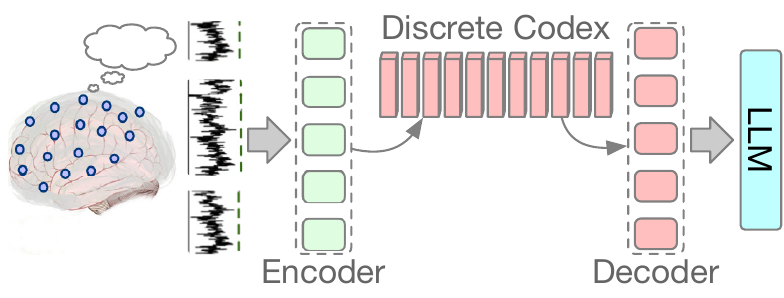}
    \end{minipage}
    \hfill
    \begin{minipage}[]{1\linewidth}
    \small  
    \centering
    \begin{tabular}{p{1cm}p{12cm}}
    \toprule
    \textbf{Ground Truth} & Bush attended \underline{\textbf{the University of} Texas \textbf{at}}  \underline{\textbf{Austin}}, \underline{where he graduated \textbf{Phi Beta Kappa}} with a Bachelor's degree in \underline{Latin \textbf{American Studies in} 1973}, taking only two and \textbf{a half years} to \underline{complete his work}, and obtaining generally \textbf{excellent grades.}\\ \midrule
    \textbf{Predict} &  was \underline{\textbf{the University of} California \textbf{at Austin}} in \underline{where he studied in \textbf{Beta Kappa}} in a degree of degree in \underline{history \textbf{American Studies in} 1975}. and a one classes \textbf{a half years} to \underline{complete the degree}. and was a \textbf{excellent grades.} \\ \bottomrule
    \end{tabular}
    \end{minipage}
    \caption{Overall illustration of translating EEG waves into text through quantised encoding.  \label{fig:coverset_ap}}
\end{figure}

In this material, we will give more technical details as well as additional experiments to support the main paper. 
The overview of the proposed framework, DeWave, is illustrated in Figure~\ref{fig:coverset_ap}. 

\section{Dataset}
\label{ap:dataset}

ZuCo stands for Zurich Cognitive Language Processing Corpus (ZuCo), a dataset that includes both raw and preprocessed eye-tracking and electroencephalography (EEG) data. The data is collected by having human subjects read given text corpora while simultaneously recording both their eye-tracking signals and EEG waves. The recording is done using the Biosemi-128 system, which, after denoising, provides 105 out of 128 channels for downstream tasks. The dataset comprises two versions: ZuCo 1.0, collected from 12 subjects, and ZuCo 2.0, collected from 18 subjects~\cite{hollenstein2018zuco,hollenstein2020zuco}.

The text corpora within the ZuCo dataset are sourced from a diverse set of textual genres, including 1) Wikipedia articles, 2) movie reviews, and 3) the BNC (British National Corpus). This diversity ensures a wide variety of syntactic structures and word frequencies. The dataset records data during two tasks: Normal Reading (NR) and Task-Specific Reading (TSR).
In our experiments, DeWave utilizes both ZuCo 1.0~\cite{hollenstein2018zuco} and 2.0~\cite{hollenstein2020zuco}. The EEG features are captured using a 128-channel system with a sampling rate of 500Hz, filtered through a frequency band ranging from 0.1Hz to 100Hz. After noise canceling, only 105 channels are deemed suitable for translation~\cite{hollenstein2018zuco}.

For word-level EEG feature translation, eye-fixation data associated with each word during reading is available in the ZuCo dataset. Following the approach similar to ~\cite{wang2022open}, we extract segments of the EEG wave according to eye fixations. Words fixated upon multiple times have their EEG fragments concatenated for processing. To process these word-level EEG features, we compute statistical results across four frequency band filters: the Theta band (5-7Hz), the Alpha band (8-13Hz), the Beta band (12-30Hz), and the Gamma band (30Hz and above)~\cite{mcfarland2006bci}. Consequently, the feature size for each word totals \(105 \times 4 \times 2 = 840\). For raw EEG waves, the signals are normalized to a range between 0 and 1 for decoding.

The dataset is split into training (80\%), development (10\%), and testing (10\%) sets, comprising 10,874, 1,387, and 1,387 unique sentences, respectively, with no overlap.
We further conducted a statistical analysis on the sentences extracted from the dataset, details of which are reported below.

\begin{table}[h]
\centering
\caption{Statistical analysis of sentences from the ZuCo dataset.}
\begin{tabularx}{0.8\textwidth}{lcc}
\toprule
\textbf{Feature} & \textbf{ZuCo 1.0 Natural Reading} & \textbf{ZuCo 2.0 Natural Reading} \\
\midrule
Sentences & 300 & 390 \\
Sent. length & 21.3 \( \pm \) 10.6 & 19.6 \( \pm \) 8.8 \\
Total words & 6386 & 6828 \\
Word length & 6.7 & 4.9 \\
\bottomrule
\end{tabularx}
\end{table}

\section{Implementation Details }
\label{ap:implementation}
We release our implementation code through GitHub to contribute to this area. 
Currently the basic code are available through an anonymous link\footnote{\href{https://github.com/duanyiqun/DeWave}{https://github.com/duanyiqun/DeWave}}
For word-level EEG features, we use the $56$ tokens each with an $840$ embedding size. 
The codex encoder for word-level features is a 6-layer transformer encoder with head number 8, hidden embedding 512.
For raw EEG waves, we clip or pad the EEG waves up to sample point 5500 with a constant value of zero, which scales up to 11 seconds according to the sampling rate of 500 Hz. 
The codex encoder for raw EEG wave features is illustrated in Section~\ref{subsec:vectorization}, where a 6-layer CNN encoder slides through the whole wave and gets the embedding sequence. A transformer layer with head number 8 and a $1\times1$ convolutional layer are combined to fuse multiple EEG channels into one embedding with size 512. 

The codex encoder shares the same structure with word-level features. 
DeWave uses a codex with size 2048 where each codex latent is an embedding with size 512. The ablation study gives a discussion about the codex size. 
All models are trained on Nvidia V100 and A100 GPUs. 
For the self-supervised decoding for raw waves, we use a learning rate of 5e-4 and a VQ coefficient of 0.25 for training 35 epochs. 
For training the codex (stage 1), DeWave uses a learning rate of 5e-4 for 35 epochs. 
For finetuning the translation (stage 2), DeWave uses a learning rate of 5e-6 for 30 epochs. 
We use the SGD as the optimizer for training all the models.

\section{Training Paradigm}
\label{ap:training}
DeWave is trained through a multi-stage process, where the training process is illustrated in Appendix algorithm~\ref{alg:algorithm}. 
Before the two-stage training, if the input of the model is the raw waves, we initialize the wave2vec model with a self-supervised pre-training described in section~\ref{subsec:vectorization}.
The self-supervised training is realized by encoding raw waves into discrete codex and reconstructing the discrete codex into original raw waves. 
The training process for self-supervised initialization utilizes the SGD algorithm with a learning rate of 0.0005 for 30 epochs with 0.1 times the learning rate decrease at epoch 20.  
In the first stage, we do not involve the language model in weight updates. 
The target of the first stage is to train a proper encoder projection $ \theta_{codx}$ and a discrete codex representation $\mathcal{C}$ for the language model. 
Intuitively, if the learning of the codex is successful, the translator could receive a better representation that is closer to the original representation, word2vec embedding. 
The training for the first stage is optimized by the SGD optimizer with a learning rate of 0.0005 for 35 epochs. 
However, the language model is trained on word tokens, which may not be perfectly suitable for brain tokens. 
In the second stage, the gradient of all weights, including language model $\theta_{BART}$ is opened to fine-tune the whole system. 
The training for the second stage is optimized by the SGD optimizer with a learning rate of $1e-6$ for 35 epochs.

\input{text/algo1.tex}

\begin{figure}[hbpt]
\centering
    \begin{minipage}[]{0.4\linewidth}
    \centering
    \includegraphics[width=5.5cm]{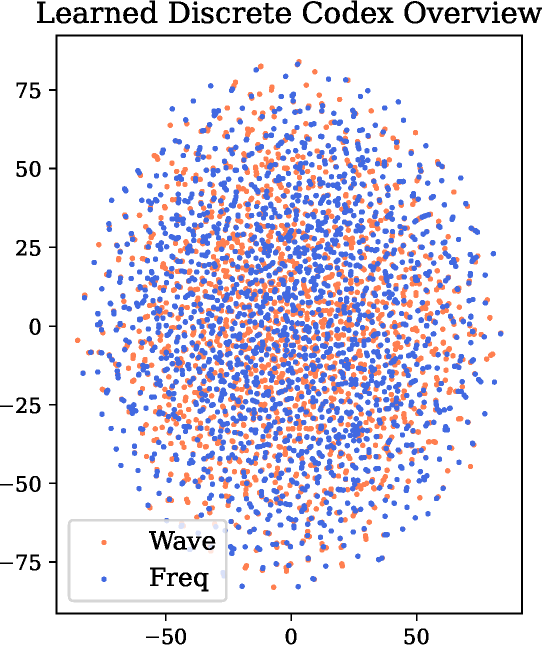}
    \end{minipage}
    \begin{minipage}[]{0.4\linewidth}
    \centering
    \includegraphics[width=5.5cm]{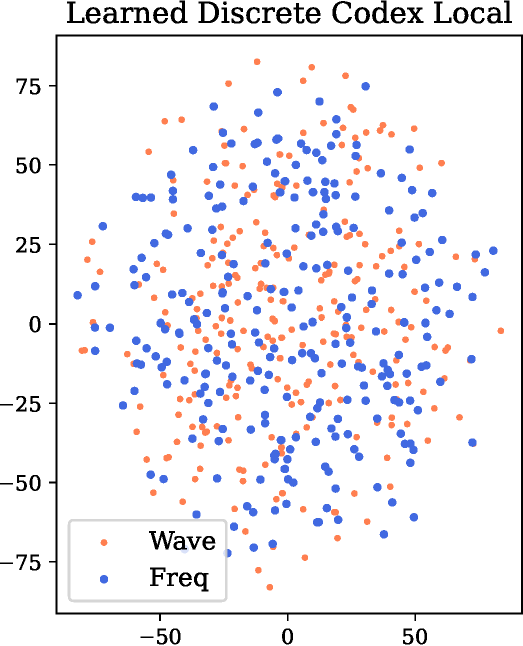}
    \end{minipage}
    \caption{Visualization of codex value distribution, where the left is the global distribution, and the right is the local distribution. \label{fig:codexdistribution}}
\end{figure}

\section{Codex Visualization}

Since the logic is to learn a discrete codex from brain dynamics, it naturally arises whether the codex value distribution between raw waves and frequency features has differences. 
In that case, we conduct additional experiments to visualize the learned codex book with T-SNE methods and report the results in Fig~\ref{fig:codexdistribution}.
Ideally, the purpose of the discrete codex is to make the language model have a better understanding of brain encoding. 
In that case, the learned codex regardless of whether it is for frequency features or raw waves, should be approximately the same as the word2vec embedding. 
In other words, the distribution of the codex should be similar. 
Fig.~\ref{fig:codexdistribution} supports our expectation, where the codex learned from frequency and raw waves have very similar distributions. 
However, the frequency codex has more coverage with corner cases and boundaries. 
We think this is rational since frequency features are naturally easier to distinguish as it has introduced manually selected features. 
The performance gap between raw waves and frequency features supports this point as well. 
Still, the similarity of the distribution illustrates the rationality of the learned codex.

\section{Motivation and Preliminary Tests with LLMs}

We provide additional insights into our experiments with larger language models. While our primary experiments utilized BART to ensure consistency in the decoder scale with prior works, it was paramount for us to ascertain that the observed improvements emanated from discrete coding and not just a more sophisticated decoder.

The potential of bridging brain activities with larger language models (LLMs) and advancing towards AGI is a significant research avenue. Recognizing this, we recently undertook an ablation study, where we replaced the BART decoder with OPT and Llama V1. Contrary to our expectations, the performance enhancement was modest. Given the vast implications of this area, we previously refrained from including these findings in the main manuscript for reasons of prudence.

Limited by our computational resources, we employed the PyTorch FSDP mode to fine-tune the OPT-1.3B and Llama-1 7B models with half-precision across three epochs. Taking cues from Mini-GPT4's method for handling visual tokens, the tokenized EEG waves were prompted into LLMs.
The performance metrics for our experiments with BART, OPT 1.3B, and Llama-1 7B decoders are tabulated below:

\begin{table}[h]
    \centering
    \caption{Performance metrics for different decoders on the ZuCo dataset.}
    \resizebox{\linewidth}{!}{
    \begin{tabularx}{1.15\linewidth}{llccccc}
        \toprule
        \textbf{Source} & \textbf{Decoder} & \textbf{BLEU-1} & \textbf{BLEU-3} & \textbf{ROUGE-R} & \textbf{ROUGE-P} & \textbf{ROUGE-F} \\
        \midrule
        Word-level features & DeWave & 41.35 & 13.92 & 28.82 & 33.71 & 30.69 \\
        Word-level features & DeWave + OPT 1.3B & 41.97 & 14.06 & 28.98 & 33.82 & 30.86 \\
        Word-level features & DeWave + Llama-1 7B & 42.84 & 15.03 & 29.42 & 35.43 & 32.05 \\
        \hdashline
        Raw Waves & DeWave & 20.51 & 5.16 & 21.18 & 29.42 & 24.27 \\
        Raw Waves & DeWave + OPT 1.3B & 21.31 & 5.84 & 22.09 & 29.94 & 25.42 \\
        Raw Waves & DeWave + Llama-1 7B & 22.05 & 6.03 & 22.45 & 30.01 & 26.08 \\
        \bottomrule
    \end{tabularx}
    }
\end{table}

From the table, it's evident that while the LLMs offer some enhancement, the gains are not as pronounced as one might expect. 
This underscores the complexity of the problem and the challenges of bridging brain activities with LLMs.
This simple experiment provides a deeper dive into our experiments with larger language models. We believe these findings offer additional perspectives and pave the way for more nuanced research in this domain.

\section{Generated Samples}

In this section, we visualize the generated decoding text results on brain waves and compare them with the ground truth in Table~\ref{tab:generation_results_1} and Table~\ref{tab:generation_results_2}. 
It suggests that the results are even better on long and simple sentences. 
For example, even for the ground truth with long and logic as below, the prediction could still match key information throughout the whole sentence.

\noindent\rule[0pt]{14cm}{0.08em}
 Ground Truth: \\
 Bush attended \underline{\textbf{the University of} Texas \textbf{at}}  \underline{\textbf{Austin}}, \underline{where he graduated \textbf{Phi Beta Kappa}} with a Bachelor's degree in \underline{Latin \textbf{American Studies in} 1973}, taking only two and \textbf{a half years} to \underline{complete his work}, and obtaining generally \textbf{excellent grades.} \\
\noindent\rule[0pt]{14cm}{0.08em}
Model Output:  \\
was \underline{\textbf{the University of} California \textbf{at Austin}} in \underline{where he studied in \textbf{Beta Kappa}} in a degree of degree in \underline{history \textbf{American Studies in} 1975}. and a one classes \textbf{a half years} to \underline{complete the degree}. and was a \textbf{excellent grades.} \\
\noindent\rule[0pt]{14cm}{0.08em}
People is feasible to guess the meaning of a human based on the translation from brain waves.
In the example above, the model recognizes through waves that \textbf{the University of} xxx at Austin. 
Although it is a factual mistake that the University of California is not at Austin, it still suggests that the model could approximately capture the semantic meaning through non-invasive brain waves.
A similar observation applies that the model recognizes that it is the \underline{xx \textbf{American Studies in} xx years} however the model predicts \underline{history \textbf{American Studies in} 1975} rather than the ground truth is \underline{Latin \textbf{American Studies in} 1973}. 
Surprisingly, even the years have correlations at this stage.

\input{text/appendix_generate.tex}

\newpage
\section{Subject Wise Evaluation}\label{sec:subjectwise_ap}

Section~\ref{subsec:metrics} introduces subject-wise metric evaluation on word-level EEG features by removing the subject to be tested from the training data, and then training the model from scratch for testing.
The results are shown in Fig~\ref{fig:crossradar} in the main paper, where different subjects share the same reading article. 
However, in supplementary details, we conduct a more detailed subject-wise evaluation in that we respectively train the model on every single subject on task v2.0 and test every subject to report the cross-subject performance. 
The single subject denotes the model only trained on a single subject on the task v2.0 dataset. 
For each subject, however, due to limited data, we add all data from task 1.0 as an assistance base. 
The results for each subject are reported below.

Here we seleceted subject YAC (Table~\ref{fig:subject_wise_YAC}), YAK (Table~\ref{fig:subject_wise_YAK}), YDG (Table~\ref{fig:subject_wise_YDG}), YFS (Table~\ref{fig:subject_wise_YFS}), YSL (Table~\ref{fig:subject_wise_YSL}), and YMD (Table~\ref{fig:subject_wise_YMD}) to report the performance. 
We visualize the metrics by clustering the same metrics value of different subjects in one radar chart. 
It is observed that the model performance might not be optimized if we train and test on the same subject. 
For example, if we train the model with task2.0 data from subject YFS and test on all subjects, the YFS subject only reaches BLUE-$\{1-4\}$ 43.32, 26.30, 14.68, and 7.62, which is lower than YDS, YAG, YRP, .. etc. which reach 43.79, 26.46, 14.94, and 8.03.
This suggests the cross-subject robustness of the proposed DeWave model. 
Also, since we use the same visualization scale for each radar chart, the area of the chart suggests the performance level. 
It is observed that if we change the training data from subject to subject, the average performance of every subject is affected by a similar trend.
We think this phenomenon is caused by the different signal-to-noise ratios. 
Some subjects might naturally have less noise interference which makes it easier for the model to learn meaningful features during the training process.

\section{Supplementatary Conclusion}
In this supplementary material, we give implementation details, training schema, and most importantly, more generated results and subject-wise evaluation of the proposed DeWave model. 
The generated results suggest a surprisingly good correlation between the model output on brain waves and the ground truth, even in long sentences with logic. 
Although there are factual and cognitive mistakes in the translation, it is still feasible to guess the meaning of a human based on the translation from brain waves.
The subject-wise evaluation suggests the DeWave model is stable across different human subjects.
Please refer to the tables attached below.

\input{text/subjcet_wise_eavaluation.tex}


%% file: text/algo1.tex
\begin{algorithm}[t]
    \caption{Training procedure}
    \label{alg:algorithm}
    \textbf{Input}: EEG $\mathcal{E}$, Vocabulary $\mathcal{V}$, Marker$ \mathcal{F}$, Target $\mathcal{W}$\\
    \textbf{Parameter}:  Codex $\mathcal{C}=\{\mathbf{c}\}$, $\theta_{codex}$, $ \theta_{BART}$, $\Theta_{wave}$ \\
    \begin{algorithmic}[1] 
        \IF{decode raw waves}
        \STATE  $\mathop{\arg\min}\limits_{\mathcal{C}, \theta_{codex}, \Theta_{wave}}L_{\rm{wave}}$ 
        \STATE Vectorize $\mathcal{X}=\Theta(\mathcal{E})$
        \ELSE
        \STATE Vectorize $\mathcal{X}=\Theta(\mathcal{E}, \mathcal{F})$
        \ENDIF
        \STATE Stage-1: Train codex
        \WHILE{Iteration steps}
        \STATE $\mathop{\arg\min}\limits_{\mathcal{C}, \theta_{codex}}$ $L_{\rm{}}(\mathcal{X})$  
        \ENDWHILE 
        \STATE Stage-2: Finetune Language Model
        \WHILE{Iteration steps}
        \STATE $\mathop{\arg\min}\limits_{\mathcal{C}, \theta_{codx}, \theta_{BART}}$ $L_{\rm{}}(\mathcal{X}) $  
        \ENDWHILE
        \STATE \textbf{return} $\mathcal{C}$, $\theta_{codex}$, $ \theta_{BART}$, $\Theta_{wave}$ 
    \end{algorithmic}
\end{algorithm}

%% file: text/appendix_generate.tex
\begin{table*}[hbpt]
    \small
    \centering
    \caption{Translation comparison between the ground truth and the prediction on brain waves with eye fixation on task v2.0 dataset.  \label{tab:generation_results_1} }
    \begin{tabular}{l p{13cm}}
    \toprule 
     \multirow{2}{*}{(1)} 
     & Ground Truth: The book was awarded the 1957 Pulitzer Prize for Biography ...\\
     \cmidrule{2-2}
     & Prediction:first is published the Pulitzer Pulitzer Prize for Literatureography ...\\
     \midrule
     
     \multirow{2}{*}{(2)} 
     & Ground Truth: Kennedy's other decorations of the Second World War include the Purple Heart, Asiatic-Pacific Campaign Medal, and the World War II Victory Medal. \\
     \cmidrule{2-2}
     & Prediction:  eth was son son were the day World War were a famous Heart and thepenatic StarAmerican,,, and the American War II Victory Medal. \\
     \midrule
     
     \multirow{2}{*}{(3)} 
     & Ground Truth: In 1958, Kennedy published the first edition of his book A Nation of Immigrants, closely following his involvement in the Displaced Persons Act and the 1957 bill to bring families together. \\
     \cmidrule{2-2}
     & Prediction:  the, the was his novel of of his autobiography, Life of Millionsigrants, which followed the experiences in the Vietnamrael Persons Movement of the Civil assassination of abolish it together. \\
    \midrule
    
     \multirow{2}{*}{(4)} 
     & Ground Truth: After World War II, Kennedy entered politics (partly to fill the void of his popular brother, Joseph P. Kennedy, Jr., on whom his family had pinned many of their hopes but who was killed in the war) ...\\
     \cmidrule{2-2}
     & Prediction: the War II, the was the asasly as avoid a void left a father father, John Kennedy. Kennedy), who.) who the he father had been the hopes the hopes). who had assassinated in the war. ... 
     \\
     \midrule
     
     \multirow{2}{*}{(5)} 
     & Ground Truth: In 1946, Representative James Michael Curley vacated his seat in an overwhelmingly Democratic district to become mayor of Boston and Kennedy ran for that seat, beating his Republican opponent by a large margin. \\
     \cmidrule{2-2}
     & Model Output: the, the John W Smithley was the seat in the unsuccessful Republican Congress of become a of New. become's for president office in which incumbent opponent opponent, a landslide margin.\\ 
     \midrule
     
     \multirow{2}{*}{(6)} 
     & Ground Truth: He was reelected twice, but had a mixed voting record, often diverging from President Harry S. Truman and the rest of the Democratic Party. \\
     \cmidrule{2-2}
     & Model Output: was a- to in in lost to less record record. and votingting from the Obama Truman. Truman's his Republican of the Republican Party. \\ 

     \midrule
     
     \multirow{2}{*}{(7)} 
     & Ground Truth: He was reelected twice, but had a mixed voting record, often diverging from President Harry S. Truman and the rest of the Democratic Party. \\
     \cmidrule{2-2}
     & Model Output: was a- to in in lost to less record record. and votingting from the Obama Truman. Truman's his Republican of the Republican Party. \\

     \midrule
     
     \multirow{2}{*}{(7)} 
     & Ground Truth:However, the U.S. Navy accepted him in September of that year.\\
     \cmidrule{2-2}
     & Model Output:  it film.S. government has the as the. that year. \\

     \midrule
     
     \multirow{2}{*}{(8)} 
     & Ground Truth: In the spring of 1941, Kennedy volunteered for the U.S. Army, but was rejected, mainly because of his troublesome back. \\
     \cmidrule{2-2}
     & Model Output:   the meantime of 2016, the was to the first.S. Army. and was discharged for and because he his age temper.  \\

     \midrule
     
     \multirow{2}{*}{(9)} 
     & Ground Truth: When Bush was seventeen, he went to Leon, Mexico, as part of his school's student exchange program. \\
     \cmidrule{2-2}
     & Model Output:  the was president, he was to aidas Nebraska, to a of a father's " exchange program.\\

     \midrule
     
     \multirow{2}{*}{(10)} 
     & Ground Truth: In November 1977 he was sent to the Venezuelan capital of Caracas, in South America, to open a new operation for the bank. \\
     \cmidrule{2-2}
     & Model Output:  the,, was born to the United prison, Manacas to where a America, to work a restaurant bank. the government.\\

     \midrule
     
     \multirow{2}{*}{(11)} 
     & Ground Truth: In 1923 he was awarded the inaugural Bôcher Memorial Prize by the American Mathematical Society. \\
     \cmidrule{2-2}
     & Model Output:   the, was born the Nobel PulitzerAFTAne Prize Medal for the French Academyical Society.  \\

     \midrule
     
     \multirow{2}{*}{(12)} 
     & Ground Truth: The mathematician Garrett Birkhoff (1911-1996) was his son. \\
     \cmidrule{2-2}
     & Model Output:  tfirst and Wkoff was1802-19) was a name. \\

    
    \bottomrule
    \end{tabular}

\end{table*}

\begin{table*}[hbpt]
    \small
    \centering
    \caption{Translation comparison between the ground truth and the prediction on brain waves with eye fixation on task v2.0 dataset. \label{tab:generation_results_2}}
    \begin{tabular}{l p{13cm}}
    \toprule 
     
     \multirow{2}{*}{(13)} 
     & Ground Truth: Jeb Bush was born in Midland, Texas, where his father was running an oil drilling company.\\
     \cmidrule{2-2}
     & Model Output:  uan Bush was a in 18way, Texas, in he father was an insurance refinery company. \\

     \midrule
     
     \multirow{2}{*}{(14)} 
     & Ground Truth:  He was noted for his lyrical playing, and performed with John Coltrane, Dexter Gordon, Hampton Hawes, Jackie McLean, and Ike and Tina Turner, among others. \\
     
     \cmidrule{2-2}
     & Model Output: was a for his "ical, style and his a a Legendtrane in who Gordon and and Fes and and GleGovern and and others Turner Tina Turner. among others. \\

    \midrule
     \multirow{2}{*}{(15)} 
     & Ground Truth: He later became an educator, teaching music theory at the University of the District of Columbia; he was also director of the District of Columbia Music Center jazz workshop band. \\
     \cmidrule{2-2}
     & Model Output:  was added a actor and and at to and the University of California Arts of Columbia. he was also a of the University's Columbia's Festival. program.. \\
     \midrule
     
     \multirow{2}{*}{(16)} 
     & Ground Truth: John Ellis "Jeb" Bush (born February 11, 1953), a Republican, is the forty-third and current Governor of Florida.\\
     \cmidrule{2-2}
     & Prediction:  nie,Johnock" Ellis (19 18 17, 18) a former, was the author-six president final governor of Texas. \\
     \midrule
     
     \multirow{2}{*}{(17)} 
     & Ground Truth:  He is a prominent member of the Bush family, the younger brother of President George W. Bush and the second son of former President George H. W. Bush and Barbara Bush. \\
     \cmidrule{2-2}
     & Prediction:   was a former member of the American family and and first brother of President Bush Bush. Bush. the father son of President President Richard H. W. Bush. his Bush. \\
    \midrule
    
     \multirow{2}{*}{(18)} 
     & Ground Truth: After earning his degree, Bush went to work in an entry level position in the international division of Texas Commerce Bank, which was run by Ben Love. ...\\
     \cmidrule{2-2}
     & Prediction: the his degree, he was on work for the office- position in the Department banking of the A.. where was later by his Carsonll ... 
     \\
     \midrule
     
     \multirow{2}{*}{(19)} 
     & Ground Truth: Following the 1980 presidential election, Bush and his family moved to Miami-Dade County, Florida. \\
     \cmidrule{2-2}
     & Model Output: the deaths election, the was his wife moved to California,Dade County, Florida,\\ 
     \midrule
     
     \multirow{2}{*}{(20)} 
     & Ground Truth: He took a job in real estate with Armando Codina, a 32-year-old Cuban immigrant and self-made American millionaire. \\
     \cmidrule{2-2}
     & Model Output:   was a liking as the estate in aando Iino in a former-year-old from immigrant from former-made millionaire businessman.\\ 

     \midrule
     
     \multirow{2}{*}{(21)} 
     & Ground Truth: [4] Situated in Liberty City, Dade County, the school is located just outside of greater Miami, in an area plagued by poverty. \\
     \cmidrule{2-2}
     & Model Output: ..]] Theuations in the,, Missouri. County, North city is a in outside of the New. Florida the area known by crime and \\

     \midrule
     
     \multirow{2}{*}{(22)} 
     & Ground Truth:The co-founder, working alongside Bush as a partner, was T. Williard Fair, a well-known local black activist and head of the Greater Miami Urban League.\\
     \cmidrule{2-2}
     & Model Output:  first-founder of John with his, a consultant, was aoniJard,banks who former-known film politiciansmith and activist of the Black Chicago NAACP League. \\

     \midrule
     
     \multirow{2}{*}{(23)} 
     & Ground Truth: Governor Buddy MacKay (55$\%$ to 45$\%$) to become governor, after courting moderate voters and Hispanics. \\
     \cmidrule{2-2}
     & Model Output:  or of RoKay ofleft) of 55$\%$) of the governor of and theting the opposition in winning. \\

     \midrule
     
     \multirow{2}{*}{(24)} 
     & Ground Truth: At the urging of his wife, Columba, a devout Mexican Catholic, the Protestant Bush became a Roman Catholic. \\
     \cmidrule{2-2}
     & Model Output:  the same of his wife, hea, he young Catholic Catholic, he actor pastorman a Catholic Catholic in  \\

     \midrule
     
     \multirow{2}{*}{(25)} 
     & Ground Truth: Bush attended the University of Texas at Austin, where he graduated Phi Beta Kappa with a Bachelor's degree in Latin American Studies in 1973, taking only two and a half years to complete his work, and obtaining generally excellent grades. \\
     \cmidrule{2-2}
     & Model Output:  was the University of California at Austin in where he studied in Beta Kappa in a degree of degree in history American Studies in 1975. and a one classes a half years to complete the degree. and was a excellent grades. \\

    
    \bottomrule
    \end{tabular}

\end{table*}

%% file: text/subjcet_wise_eavaluation.tex
\begin{table*}
    \begin{minipage}[]{0.23\linewidth}
    \includegraphics[width=4cm]{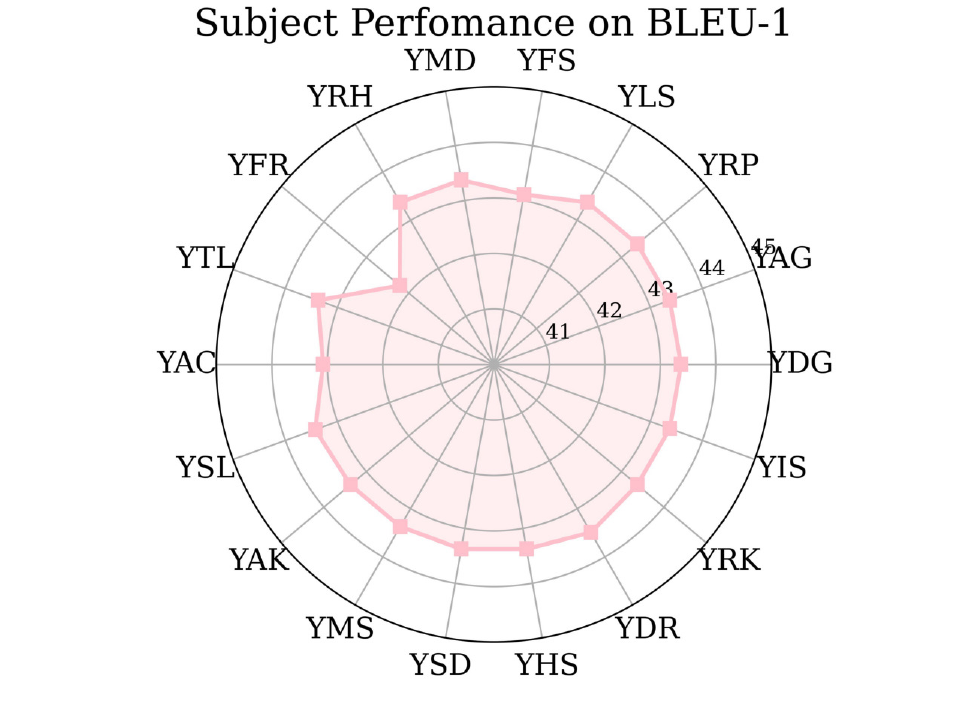}
    \end{minipage}
    \begin{minipage}[]{0.23\linewidth}
    \includegraphics[width=4cm]{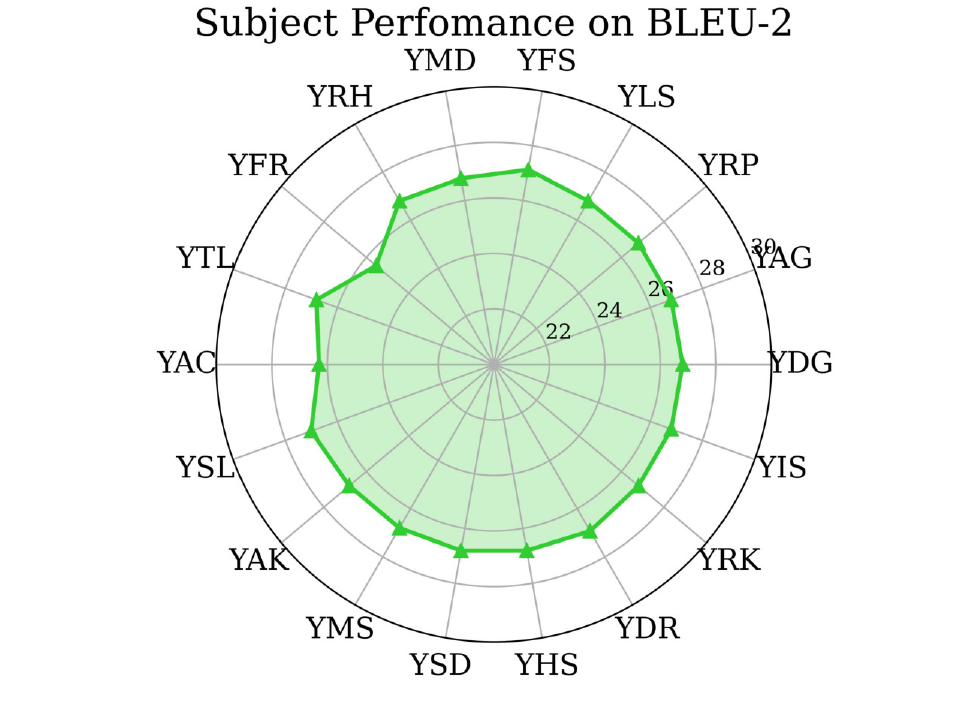}
    \end{minipage}
    \begin{minipage}[]{0.23\linewidth}
    \includegraphics[width=4cm]{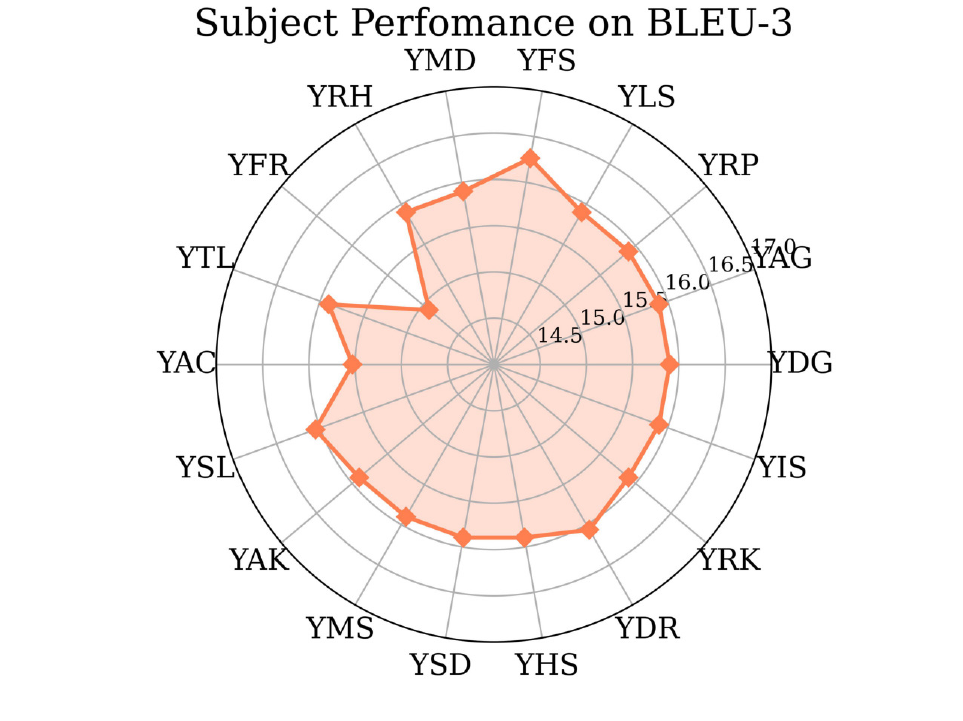}
    \end{minipage}
    \begin{minipage}[]{0.23\linewidth}
    \includegraphics[width=4cm]{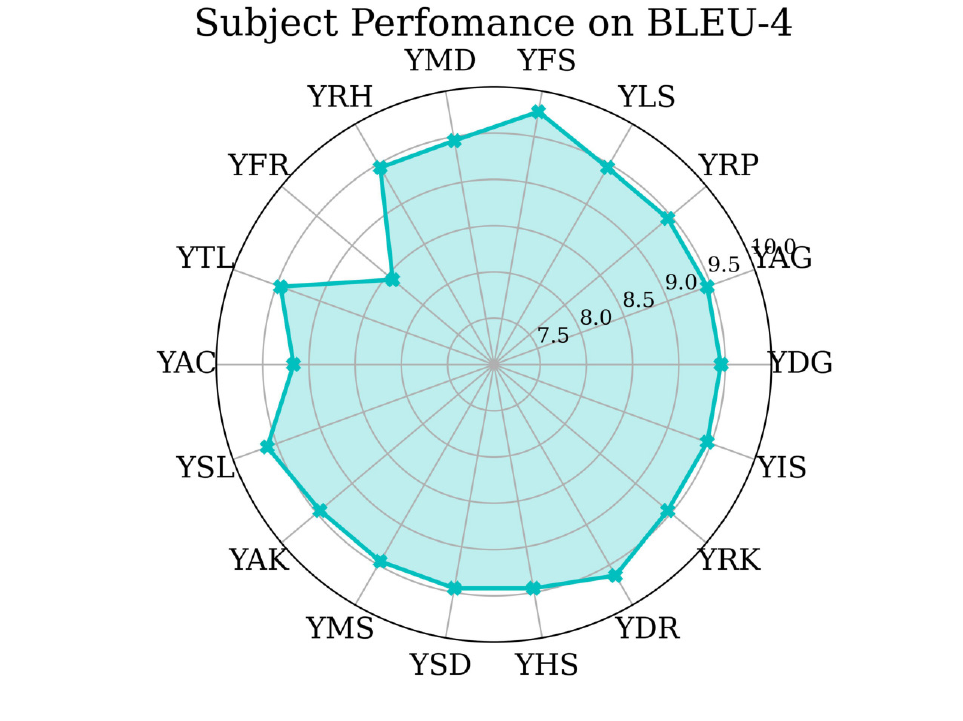}
    \end{minipage}
    \begin{minipage}[]{0.245\linewidth}
    \includegraphics[width=4cm]{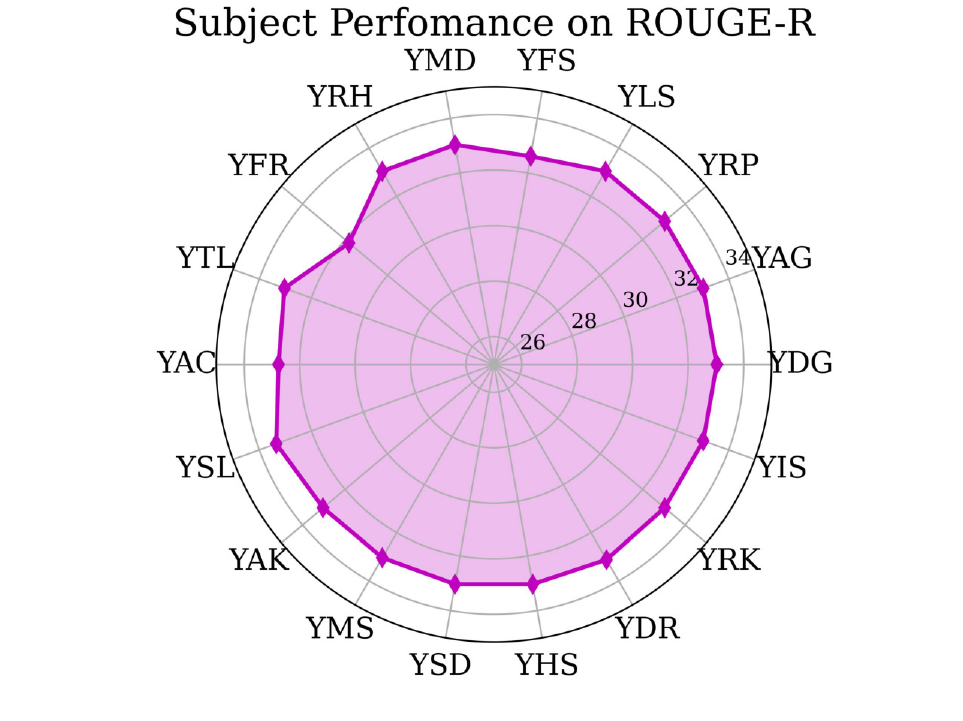}
    \end{minipage}
    \begin{minipage}[]{0.25\linewidth}
    \includegraphics[width=4cm]{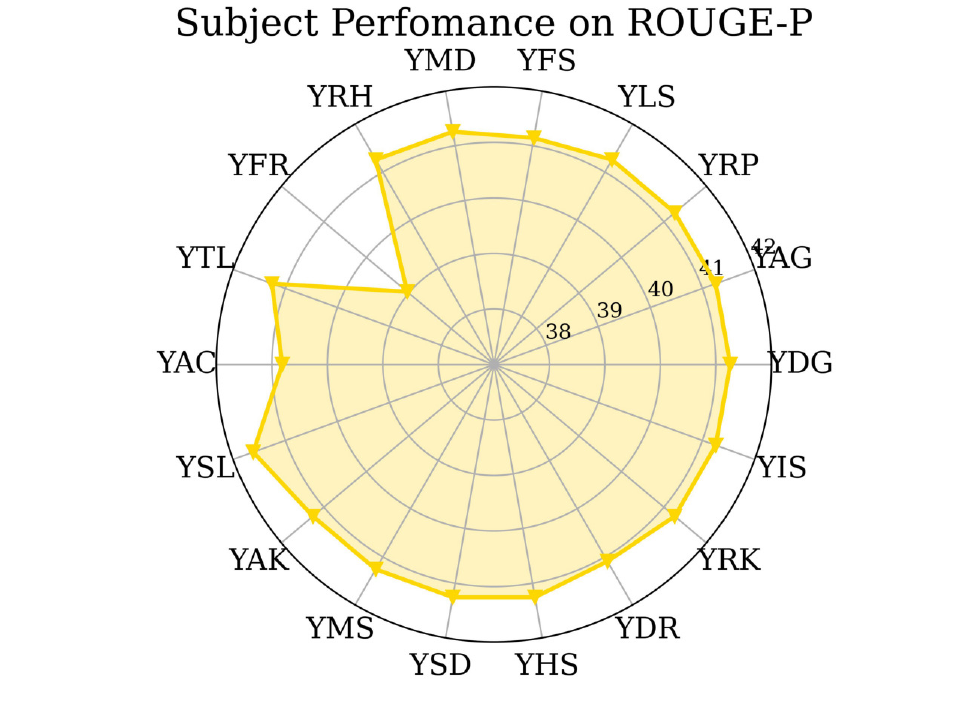}
    \end{minipage}
    \begin{minipage}[]{0.25\linewidth}
    \includegraphics[width=4cm]{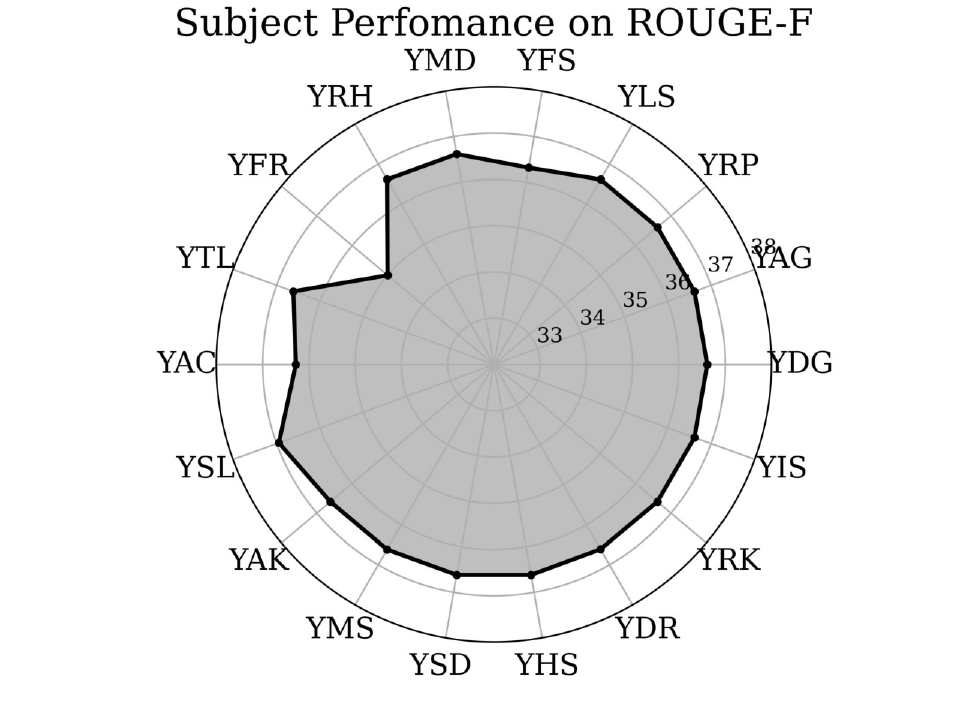}
    \end{minipage}
    \begin{minipage}[]{0.25\linewidth}
    \end{minipage}

\begin{minipage}[]{\linewidth}
\footnotesize
\resizebox{\columnwidth}{!}{
\begin{tabular}{p{1.5cm}p{0.5cm}p{0.5cm}p{0.5cm}p{0.5cm}p{0.5cm}p{0.5cm}p{0.5cm}p{0.5cm}p{0.5cm}p{0.5cm}p{0.5cm}p{0.5cm}p{0.5cm}p{0.5cm}p{0.5cm}p{0.5cm}p{0.5cm}p{0.5cm}}
\toprule
{Subject} & YDG & YAG & YRP & YLS & YFS & YMD & YRH & YFR & YTL & YAC & YSL & YAK & YMS & YSD & YHS & YDR & YRK & YIS \\ 
\midrule
{BLEU-1} & 43.14 & 43.14 & 43.14 & 43.14 & 42.88 & 43.14 & 43.14 & 41.98 & 43.14 & 42.85 & 43.20 & 43.14 & 43.14 & 43.14 & 43.14 & 43.27 & 43.14 & 43.14 \\
{BLEU-2} & 26.58 & 26.58 & 26.58 & 26.58 & 26.89 & 26.58 & 26.58 & 25.31 & 26.58 & 26.07 & 26.78 & 26.58 & 26.58 & 26.58 & 26.58 & 26.71 & 26.58 & 26.58 \\
{BLEU-3} & 15.67 & 15.67 & 15.67 & 15.67 & 16.03 & 15.67 & 15.67 & 14.68 & 15.67 & 15.30 & 15.82 & 15.67 & 15.67 & 15.67 & 15.67 & 15.83 & 15.67 & 15.67 \\
{BLEU-4} & 9.23 & 9.23 & 9.23 & 9.23 & 9.54 & 9.23 & 9.23 & 8.20 & 9.23 & 8.94 & 9.38 & 9.23 & 9.23 & 9.23 & 9.23 & 9.40 & 9.23 & 9.23 \\
{ROUGE-R} & 32.81 & 32.81 & 32.81 & 32.81 & 32.38 & 32.81 & 32.81 & 31.58 & 32.81 & 32.53 & 33.12 & 32.81 & 32.81 & 32.81 & 32.81 & 32.89 & 32.81 & 32.81 \\
{ROUGE-P} & 41.03 & 41.03 & 41.03 & 41.03 & 40.91 & 41.03 & 41.03 & 38.81 & 41.03 & 40.58 & 41.39 & 41.03 & 41.03 & 41.03 & 41.03 & 40.86 & 41.03 & 41.03 \\
{ROUGE-F} & 36.39 & 36.39 & 36.39 & 36.39 & 36.09 & 36.39 & 36.39 & 34.77 & 36.39 & 36.05 & 36.72 & 36.39 & 36.39 & 36.39 & 36.39 & 36.38 & 36.39 & 36.39 \\
\bottomrule
\end{tabular}}
\end{minipage}

\caption{Subject-wise evaluation results on a model trained with subject \textbf{YAC}, where the radar chart suggests the performance variance on different subjects on each metric. \label{fig:subject_wise_YAC}}
\end{table*}

\begin{table*}
    \begin{minipage}[]{0.23\linewidth}
    \includegraphics[width=4cm]{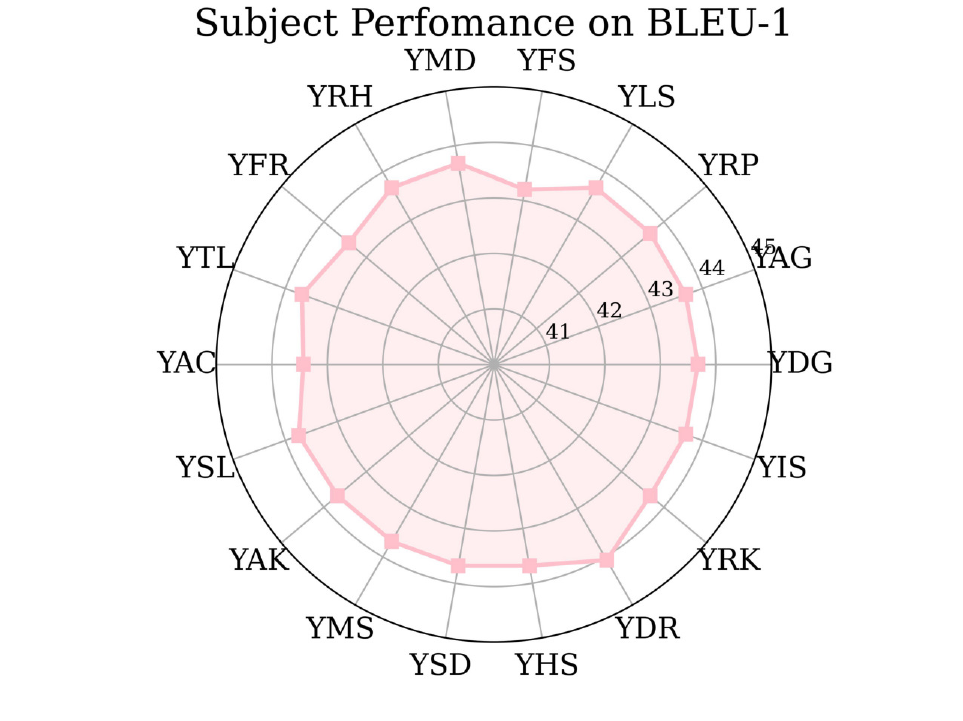}
    \end{minipage}
    \begin{minipage}[]{0.23\linewidth}
    \includegraphics[width=4cm]{image/yak/BLEU-2-eps-converted-to.pdf}
    \end{minipage}
    \begin{minipage}[]{0.23\linewidth}
    \includegraphics[width=4cm]{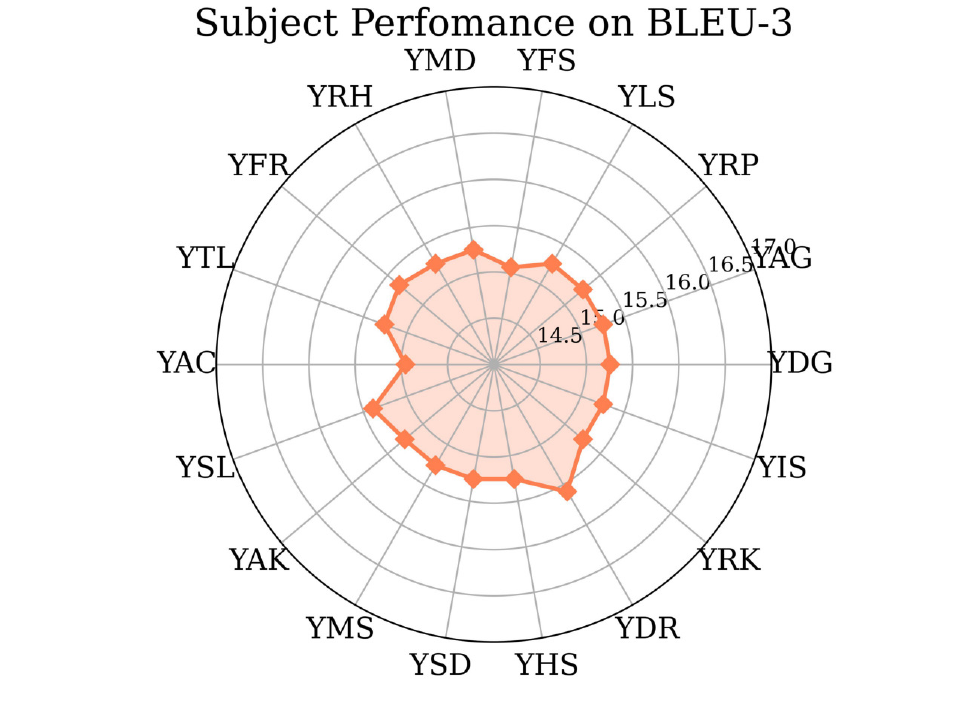}
    \end{minipage}
    \begin{minipage}[]{0.23\linewidth}
    \includegraphics[width=4cm]{image/yak/BLEU-4-eps-converted-to.pdf}
    \end{minipage}
    \begin{minipage}[]{0.245\linewidth}
    \includegraphics[width=4cm]{image/yak/ROUGE-R-eps-converted-to.pdf}
    \end{minipage}
    \begin{minipage}[]{0.25\linewidth}
    \includegraphics[width=4cm]{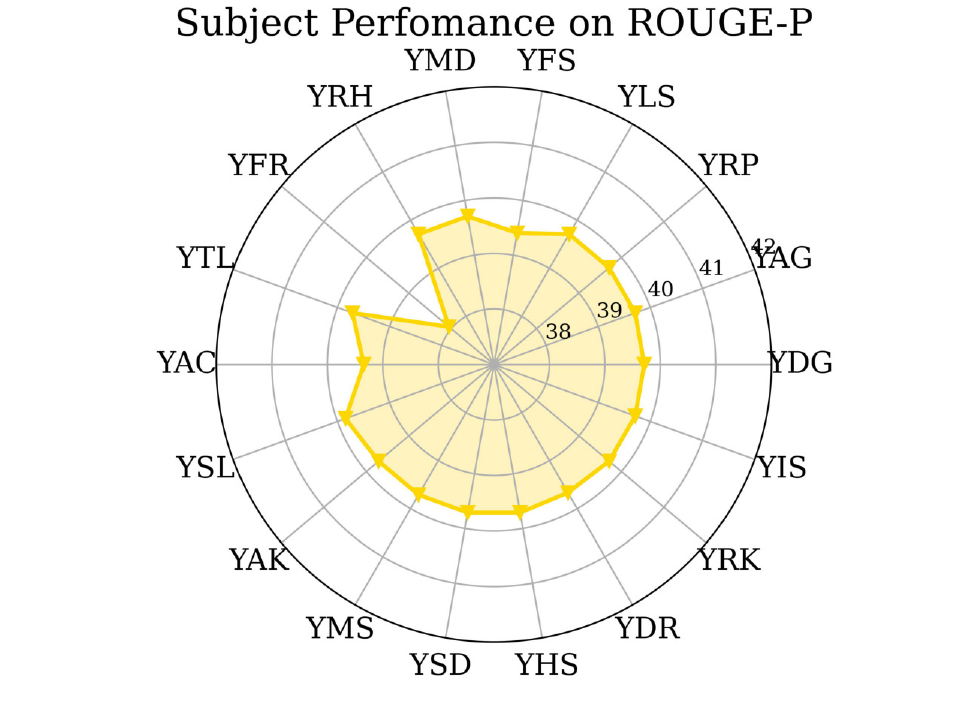}
    \end{minipage}
    \begin{minipage}[]{0.25\linewidth}
    \includegraphics[width=4cm]{image/yak/ROUGE-F-eps-converted-to.pdf}
    \end{minipage}
    \begin{minipage}[]{0.25\linewidth}
    \end{minipage}
\quad
\quad
\quad

\begin{minipage}[]{\linewidth}
\footnotesize
\resizebox{\columnwidth}{!}{
\begin{tabular}{p{1.5cm}p{0.5cm}p{0.5cm}p{0.5cm}p{0.5cm}p{0.5cm}p{0.5cm}p{0.5cm}p{0.5cm}p{0.5cm}p{0.5cm}p{0.5cm}p{0.5cm}p{0.5cm}p{0.5cm}p{0.5cm}p{0.5cm}p{0.5cm}p{0.5cm}}
\toprule
{Subject} & YDG & YAG & YRP & YLS & YFS & YMD & YRH & YFR & YTL & YAC & YSL & YAK & YMS & YSD & YHS & YDR & YRK & YIS \\ 
\midrule
{BLEU-1} & 43.68 & 43.68 & 43.68 & 43.68 & 43.20 & 43.68 & 43.68 & 43.41 & 43.68 & 43.43 & 43.74 & 43.68 & 43.68 & 43.68 & 43.68 & 44.07 & 43.68 & 43.68 \\
{BLEU-2}  & 26.34 & 26.34 & 26.34 & 26.34 & 26.26 & 26.34 & 26.34 & 25.90 & 26.34 & 25.80 & 26.53 & 26.34 & 26.34 & 26.34 & 26.34 & 26.62 & 26.34 & 26.34 \\
{BLEU-3} & 15.26 & 15.26 & 15.26 & 15.26 & 15.07 & 15.26 & 15.26 & 15.34 & 15.26 & 14.95 & 15.39 & 15.26 & 15.26 & 15.26 & 15.26 & 15.59 & 15.26 & 15.26 \\
{BLEU-4} & 9.02 & 9.02 & 9.02 & 9.02 & 8.77 & 9.02 & 9.02 & 9.01 & 9.02 & 8.75 & 9.16 & 9.02 & 9.02 & 9.02 & 9.02 & 9.27 & 9.02 & 9.02 \\
{ROUGE-R} & 32.50 & 32.50 & 32.50 & 32.50 & 32.03 & 32.50 & 32.50 & 31.68 & 32.50 & 32.02 & 32.66 & 32.50 & 32.50 & 32.50 & 32.50 & 32.56 & 32.50 & 32.50 \\
{ROUGE-P} & 39.71 & 39.71 & 39.71 & 39.71 & 39.41 & 39.71 & 39.71 & 38.05 & 39.71 & 39.34 & 39.84 & 39.71 & 39.71 & 39.71 & 39.71 & 39.67 & 39.71 & 39.71 \\
{ROUGE-F} & 35.67 & 35.67 & 35.67 & 35.67 & 35.26 & 35.67 & 35.67 & 34.51 & 35.67 & 35.24 & 35.82 & 35.67 & 35.67 & 35.67 & 35.67 & 35.69 & 35.67 & 35.67  \\
\bottomrule
\end{tabular}}
\end{minipage}

\caption{Subject-wise evaluation results on a model trained with subject \textbf{YAK}, where the radar chart suggests the performance variance on different subjects on each metric. \label{fig:subject_wise_YAK}}
\end{table*}

\begin{table*}
    \begin{minipage}[]{0.23\linewidth}
    \includegraphics[width=4cm]{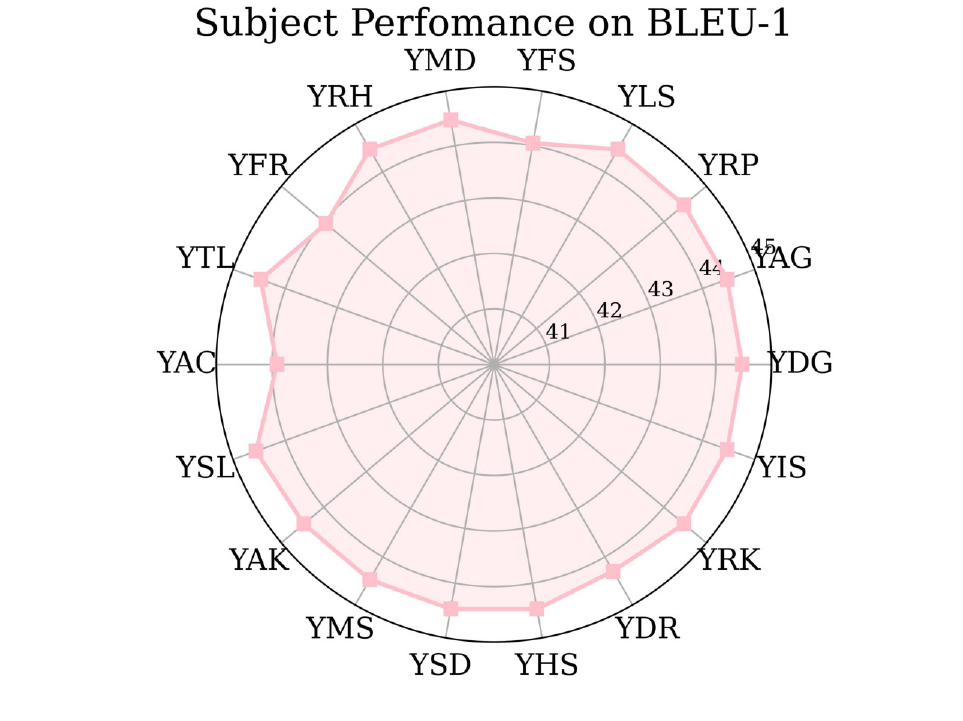}
    \end{minipage}
    \begin{minipage}[]{0.23\linewidth}
    \includegraphics[width=4cm]{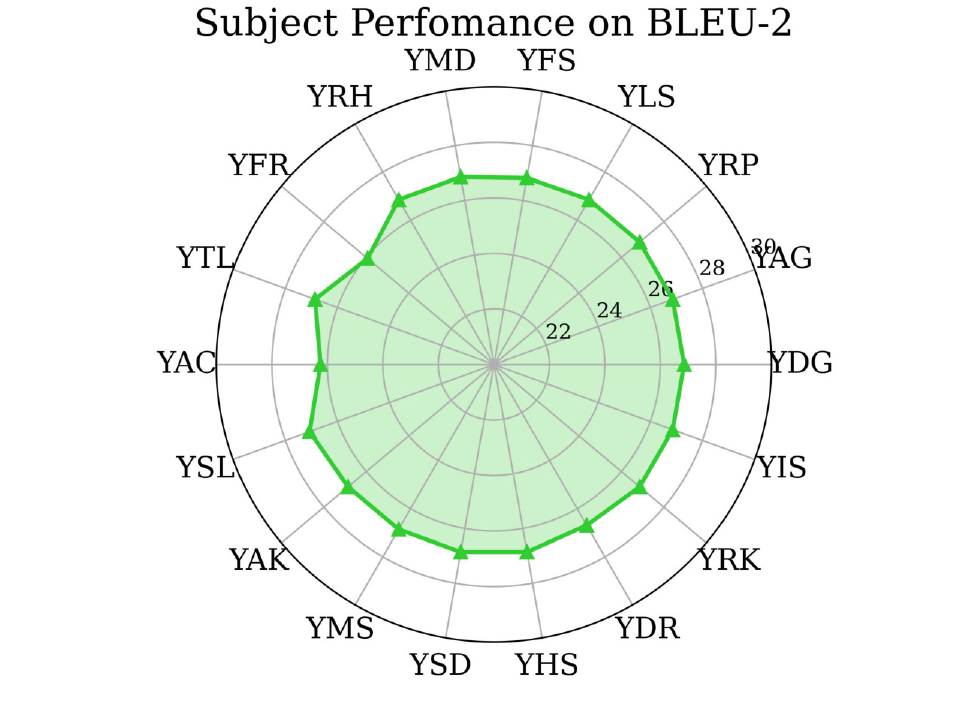}
    \end{minipage}
    \begin{minipage}[]{0.23\linewidth}
    \includegraphics[width=4cm]{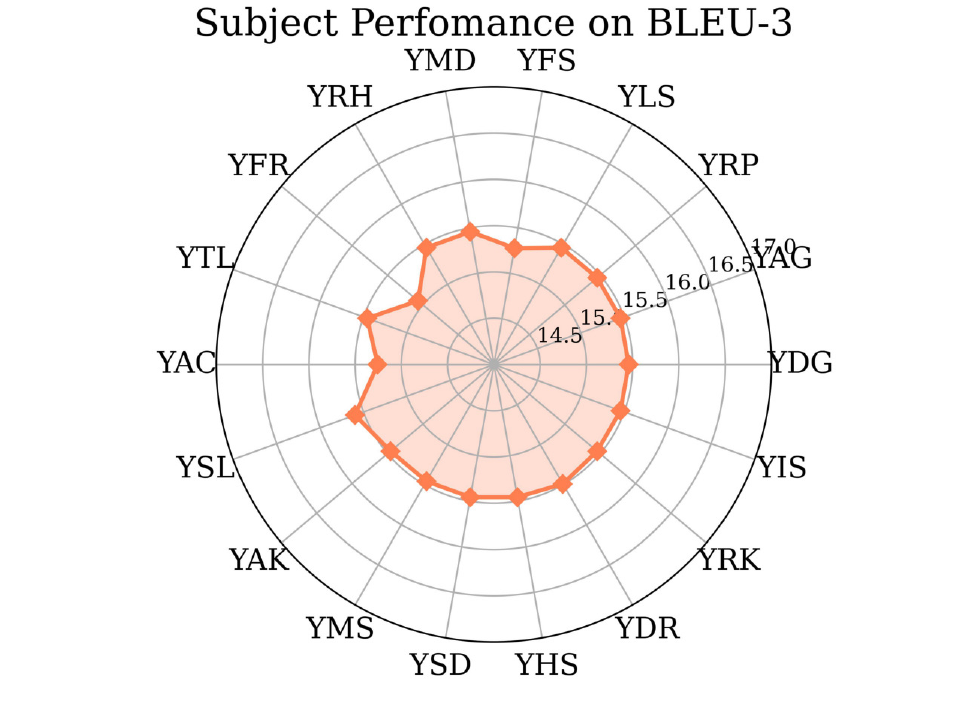}
    \end{minipage}
    \begin{minipage}[]{0.23\linewidth}
    \includegraphics[width=4cm]{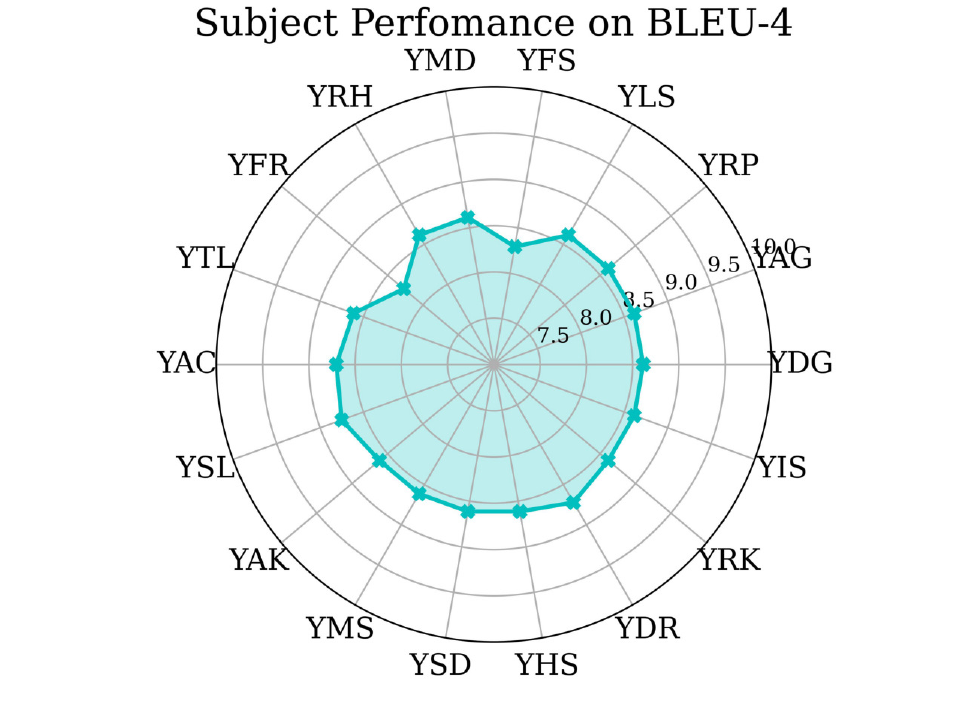}
    \end{minipage}
    \begin{minipage}[]{0.245\linewidth}
    \includegraphics[width=4cm]{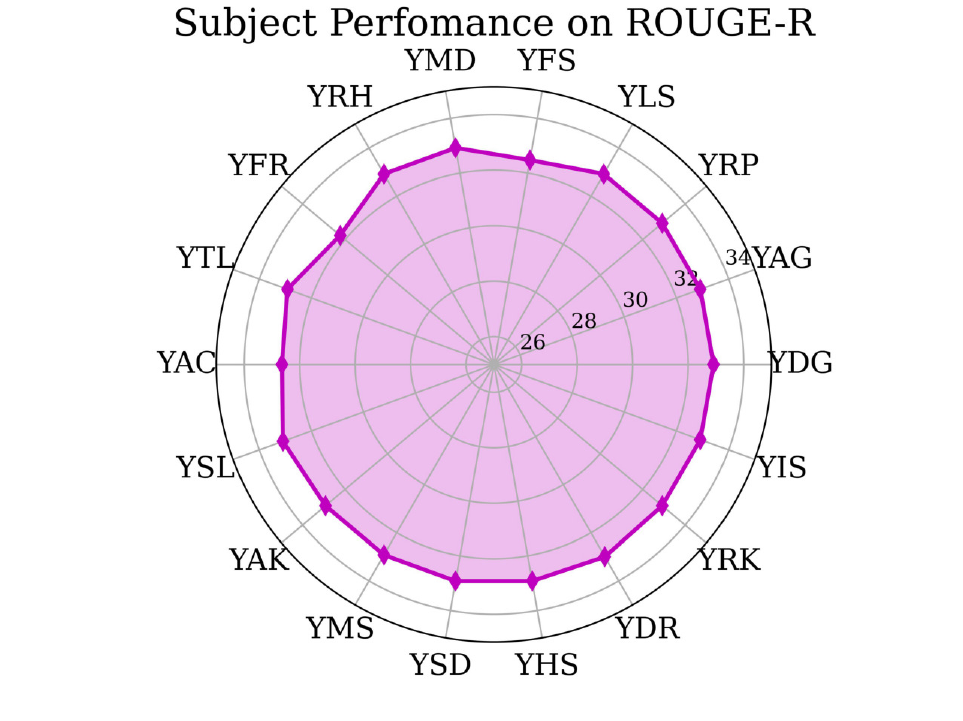}
    \end{minipage}
    \begin{minipage}[]{0.25\linewidth}
    \includegraphics[width=4cm]{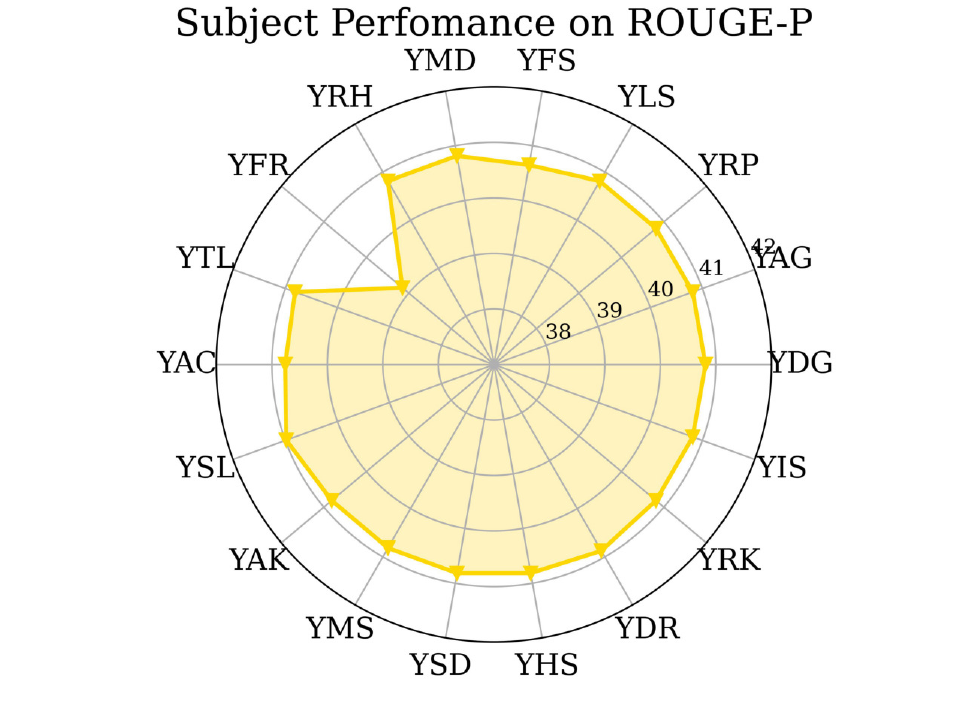}
    \end{minipage}
    \begin{minipage}[]{0.25\linewidth}
    \includegraphics[width=4cm]{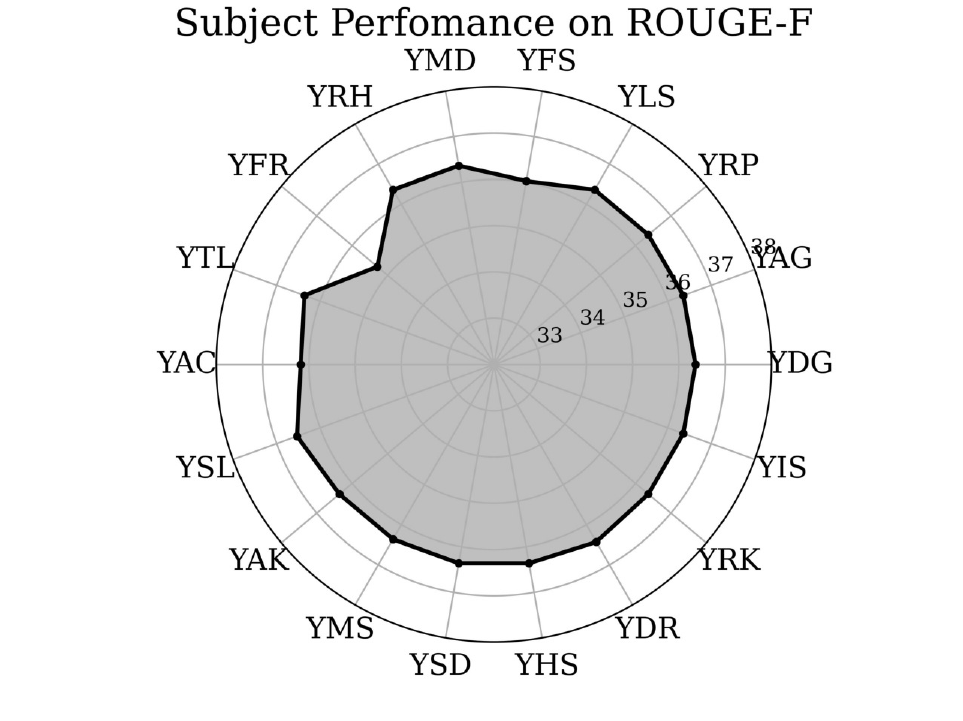}
    \end{minipage}
    \begin{minipage}[]{0.25\linewidth}
    \end{minipage}
\quad
\quad
\quad

\begin{minipage}[]{1\linewidth}
\footnotesize
\resizebox{\columnwidth}{!}{
\begin{tabular}{p{1.5cm}p{0.5cm}p{0.5cm}p{0.5cm}p{0.5cm}p{0.5cm}p{0.5cm}p{0.5cm}p{0.5cm}p{0.5cm}p{0.5cm}p{0.5cm}p{0.5cm}p{0.5cm}p{0.5cm}p{0.5cm}p{0.5cm}p{0.5cm}p{0.5cm}}
\toprule
{Subject} & YDG & YAG & YRP & YLS & YFS & YMD & YRH & YFR & YTL & YAC & YSL & YAK & YMS & YSD & YHS & YDR & YRK & YIS \\ 
\midrule
{BLEU-1} & 44.47 & 44.47 & 44.47 & 44.47 & 44.05 & 44.47 & 44.47 & 43.95 & 44.47 & 43.91 & 44.56 & 44.47 & 44.47 & 44.47 & 44.47 & 44.30 & 44.47 & 44.47 \\
{BLEU-2}  & 26.86 & 26.86 & 26.86 & 26.86 & 26.82 & 26.86 & 26.86 & 25.94 & 26.86 & 26.25 & 27.07 & 26.86 & 26.86 & 26.86 & 26.86 & 26.69 & 26.86 & 26.86 \\
{BLEU-3} & 15.46 & 15.46 & 15.46 & 15.46 & 15.28 & 15.46 & 15.46 & 15.07 & 15.46 & 15.26 & 15.60 & 15.46 & 15.46 & 15.46 & 15.46 & 15.49 & 15.46 & 15.46 \\
{BLEU-4} & 8.61 & 8.61 & 8.61 & 8.61 & 8.29 & 8.61 & 8.61 & 8.27 & 8.61 & 8.70 & 8.75 & 8.61 & 8.61 & 8.61 & 8.61 & 8.72 & 8.61 & 8.61 \\
{ROUGE-R}& 32.92 & 32.92 & 32.92 & 32.92 & 32.47 & 32.92 & 32.92 & 32.23 & 32.92 & 32.64 & 33.09 & 32.92 & 32.92 & 32.92 & 32.92 & 33.00 & 32.92 & 32.92 \\
{ROUGE-P}  & 40.82 & 40.82 & 40.82 & 40.82 & 40.65 & 40.82 & 40.82 & 39.15 & 40.82 & 40.76 & 40.98 & 40.82 & 40.82 & 40.82 & 40.82 & 40.87 & 40.82 & 40.82 \\
{ROUGE-F} & 36.36 & 36.36 & 36.36 & 36.36 & 36.03 & 36.36 & 36.36 & 35.29 & 36.36 & 36.18 & 36.53 & 36.36 & 36.36 & 36.36 & 36.36 & 36.43 & 36.36 & 36.36 \\
\bottomrule
\end{tabular}}
\end{minipage}

\caption{Subject-wise evaluation results on a model trained with subject \textbf{YDG}, where the radar chart suggests the performance variance on different subjects on each metric. \label{fig:subject_wise_YDG}}
\end{table*}

\begin{table*}
    \begin{minipage}[]{0.23\linewidth}
    \includegraphics[width=4cm]{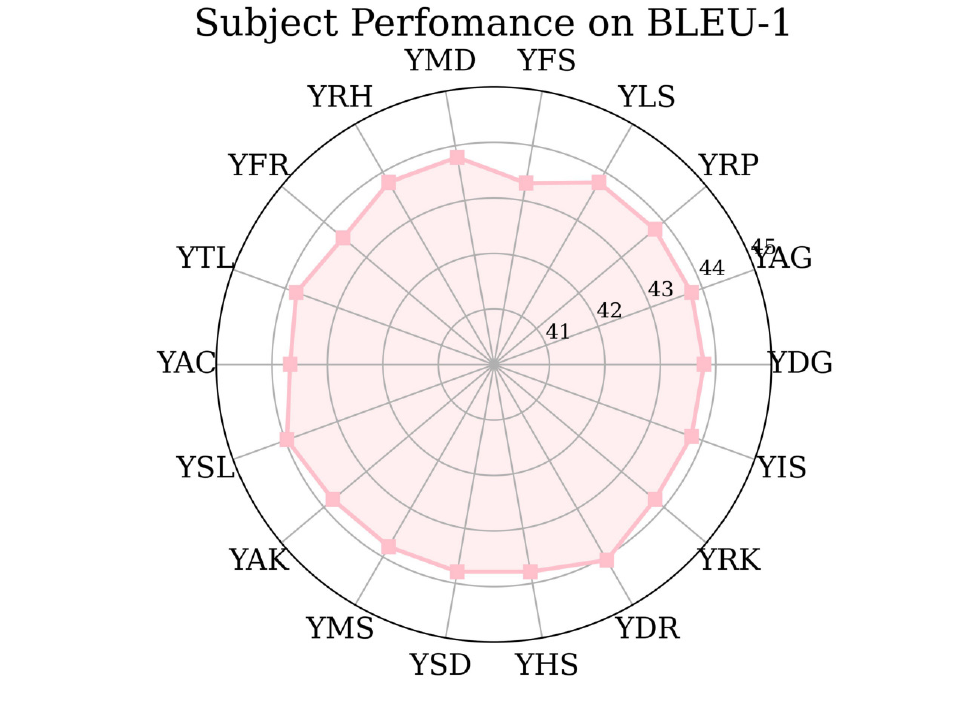}
    \end{minipage}
    \begin{minipage}[]{0.23\linewidth}
    \includegraphics[width=4cm]{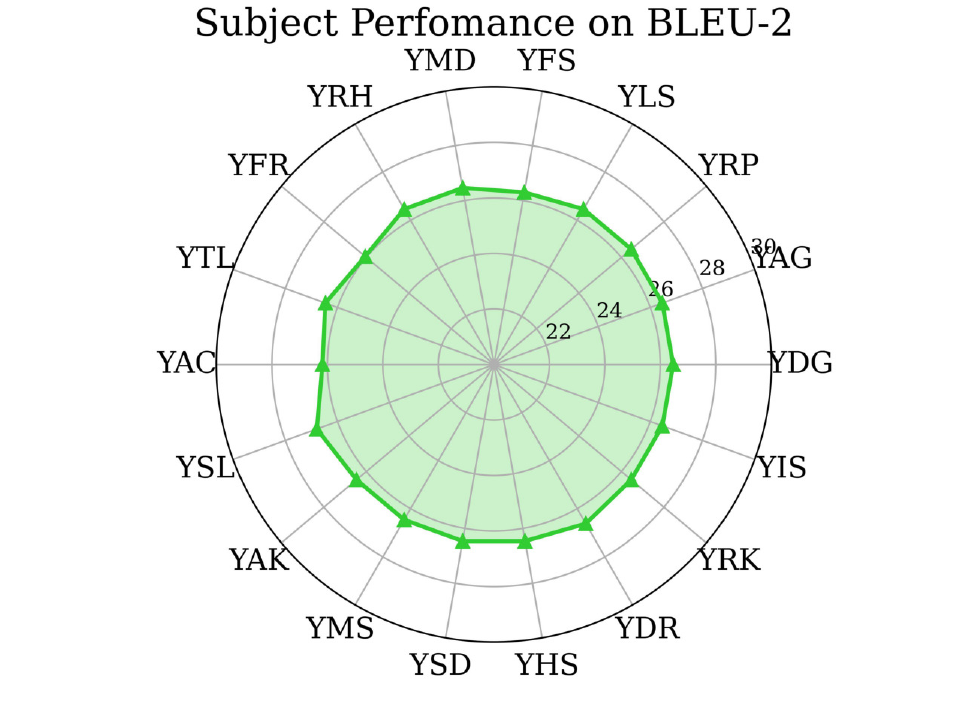}
    \end{minipage}
    \begin{minipage}[]{0.23\linewidth}
    \includegraphics[width=4cm]{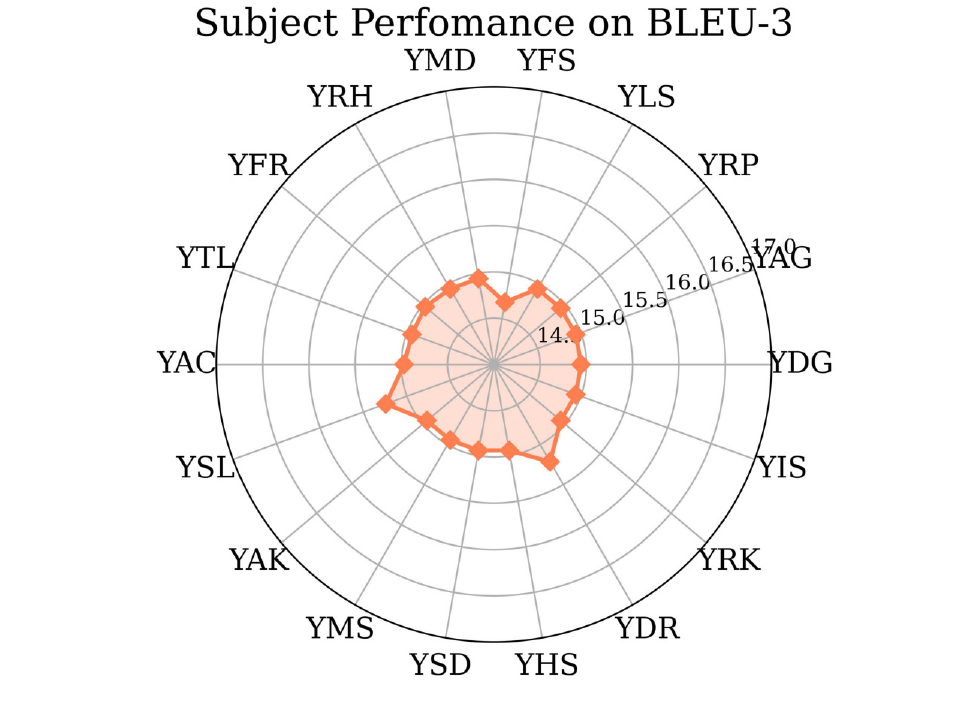}
    \end{minipage}
    \begin{minipage}[]{0.23\linewidth}
    \includegraphics[width=4cm]{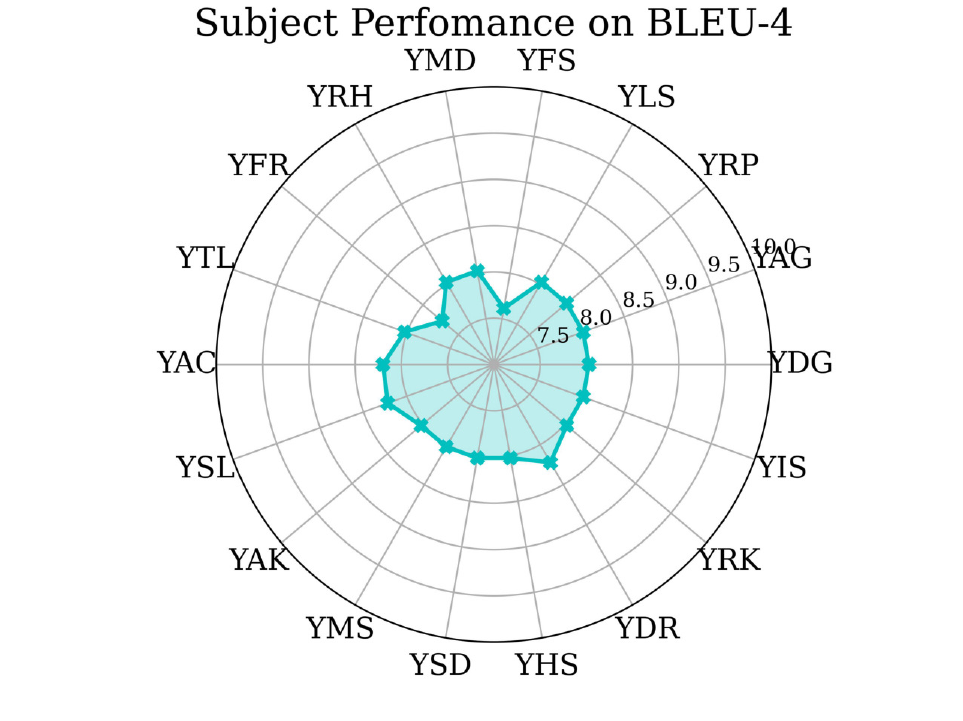}
    \end{minipage}
    \begin{minipage}[]{0.245\linewidth}
    \includegraphics[width=4cm]{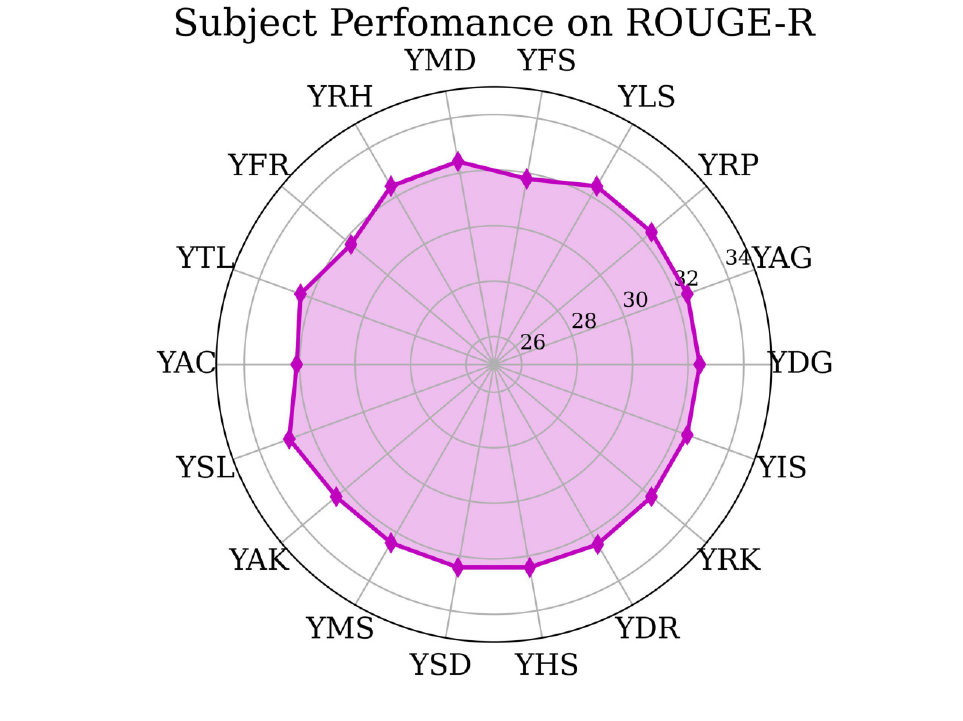}
    \end{minipage}
    \begin{minipage}[]{0.25\linewidth}
    \includegraphics[width=4cm]{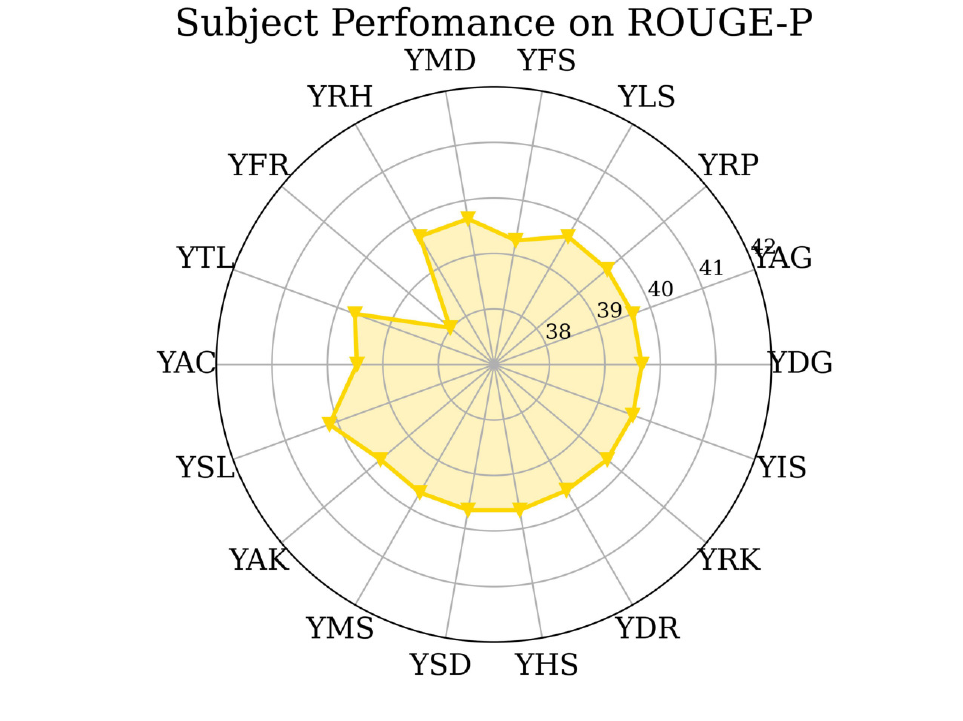}
    \end{minipage}
    \begin{minipage}[]{0.25\linewidth}
    \includegraphics[width=4cm]{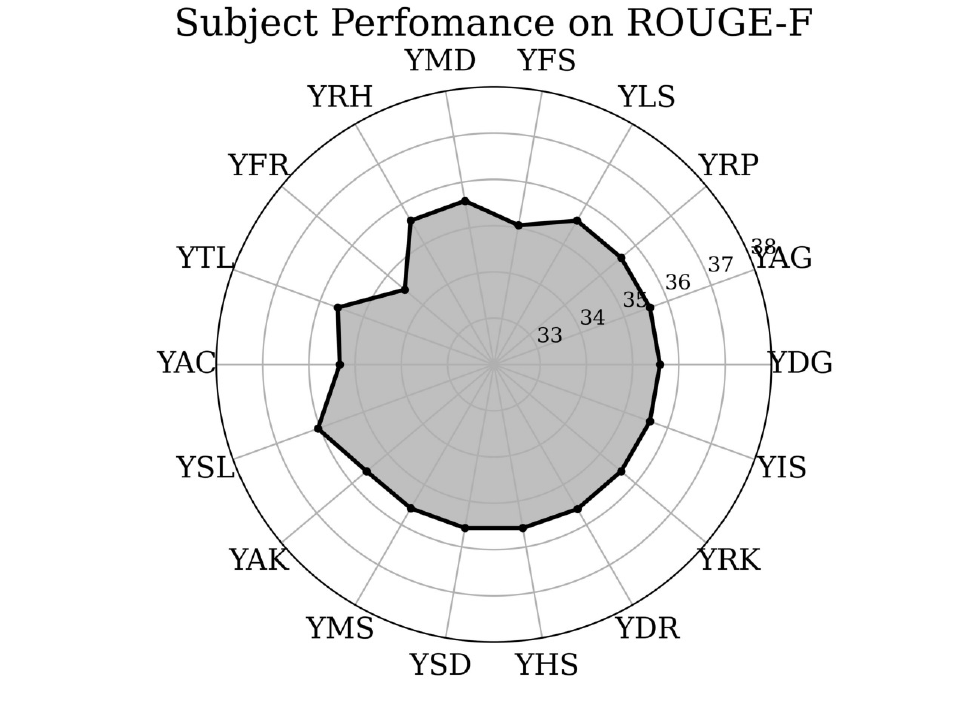}
    \end{minipage}
    \begin{minipage}[]{0.25\linewidth}
    \end{minipage}
\quad
\quad
\quad

\begin{minipage}[]{\linewidth}
\footnotesize
\resizebox{\columnwidth}{!}{
\begin{tabular}{p{1.5cm}p{0.5cm}p{0.5cm}p{0.5cm}p{0.5cm}p{0.5cm}p{0.5cm}p{0.5cm}p{0.5cm}p{0.5cm}p{0.5cm}p{0.5cm}p{0.5cm}p{0.5cm}p{0.5cm}p{0.5cm}p{0.5cm}p{0.5cm}p{0.5cm}}
\toprule
{Subject} & YDG & YAG & YRP & YLS & YFS & YMD & YRH & YFR & YTL & YAC & YSL & YAK & YMS & YSD & YHS & YDR & YRK & YIS \\ 
\midrule
{BLEU-1} & 43.79 & 43.79 & 43.79 & 43.79 & 43.32 & 43.79 & 43.79 & 43.55 & 43.79 & 43.67 & 43.97 & 43.79 & 43.79 & 43.79 & 43.79 & 44.07 & 43.79 & 43.79 \\
{BLEU-2}   & 26.46 & 26.46 & 26.46 & 26.46 & 26.30 & 26.46 & 26.46 & 26.06 & 26.46 & 26.18 & 26.79 & 26.46 & 26.46 & 26.46 & 26.46 & 26.62 & 26.46 & 26.46 \\
{BLEU-3}& 14.94 & 14.94 & 14.94 & 14.94 & 14.68 & 14.94 & 14.94 & 14.97 & 14.94 & 14.97 & 15.24 & 14.94 & 14.94 & 14.94 & 14.94 & 15.21 & 14.94 & 14.94 \\
{BLEU-4} & 8.03 & 8.03 & 8.03 & 8.03 & 7.62 & 8.03 & 8.03 & 7.73 & 8.03 & 8.20 & 8.22 & 8.03 & 8.03 & 8.03 & 8.03 & 8.22 & 8.03 & 8.03 \\
{ROUGE-R}& 32.42 & 32.42 & 32.42 & 32.42 & 31.79 & 32.42 & 32.42 & 31.72 & 32.42 & 32.11 & 32.85 & 32.42 & 32.42 & 32.42 & 32.42 & 32.48 & 32.42 & 32.42 \\
{ROUGE-P}  & 39.67 & 39.67 & 39.67 & 39.67 & 39.27 & 39.67 & 39.67 & 38.03 & 39.67 & 39.47 & 40.15 & 39.67 & 39.67 & 39.67 & 39.67 & 39.62 & 39.67 & 39.67 \\
{ROUGE-F} & 35.59 & 35.59 & 35.59 & 35.59 & 35.06 & 35.59 & 35.59 & 34.52 & 35.59 & 35.33 & 36.05 & 35.59 & 35.59 & 35.59 & 35.59 & 35.61 & 35.59 & 35.59\\
\bottomrule
\end{tabular}}
\end{minipage}

\caption{Subject-wise evaluation results on a model trained with subject \textbf{YFS}, where the radar chart suggests the performance variance on different subjects on each metric. \label{fig:subject_wise_YFS}}
\end{table*}

\begin{table*}
    \begin{minipage}[]{0.23\linewidth}
    \includegraphics[width=4cm]{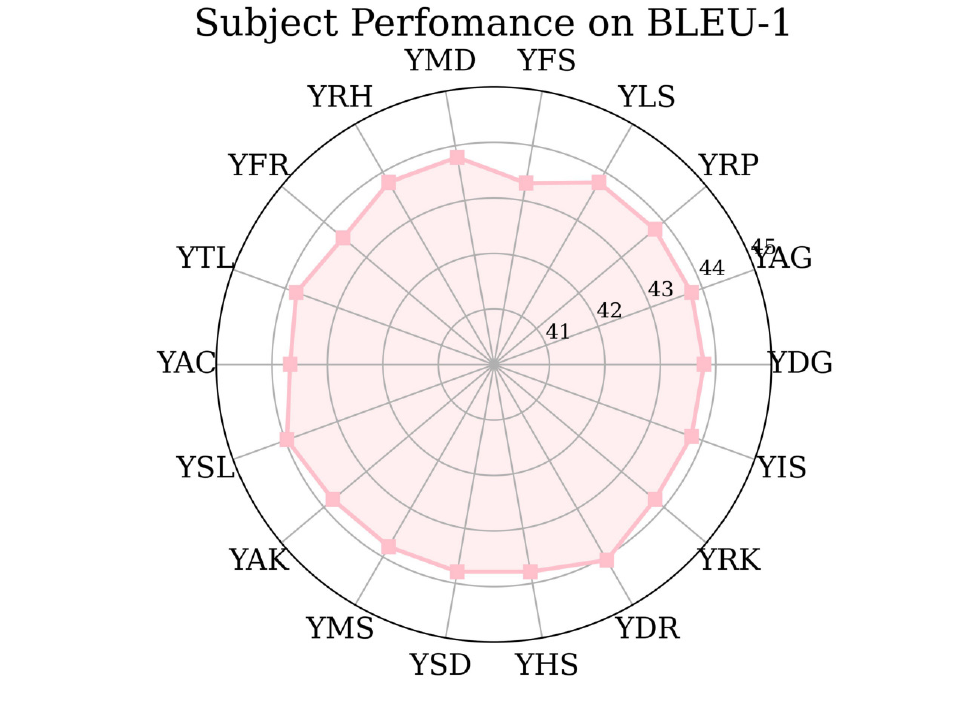}
    \end{minipage}
    \begin{minipage}[]{0.23\linewidth}
    \includegraphics[width=4cm]{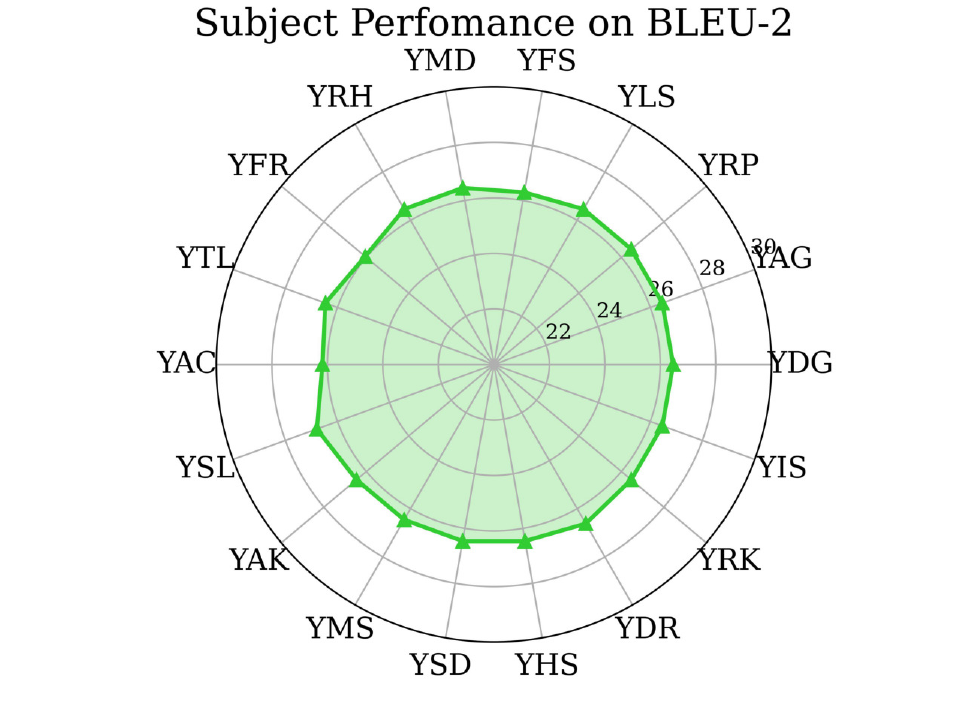}
    \end{minipage}
    \begin{minipage}[]{0.23\linewidth}
    \includegraphics[width=4cm]{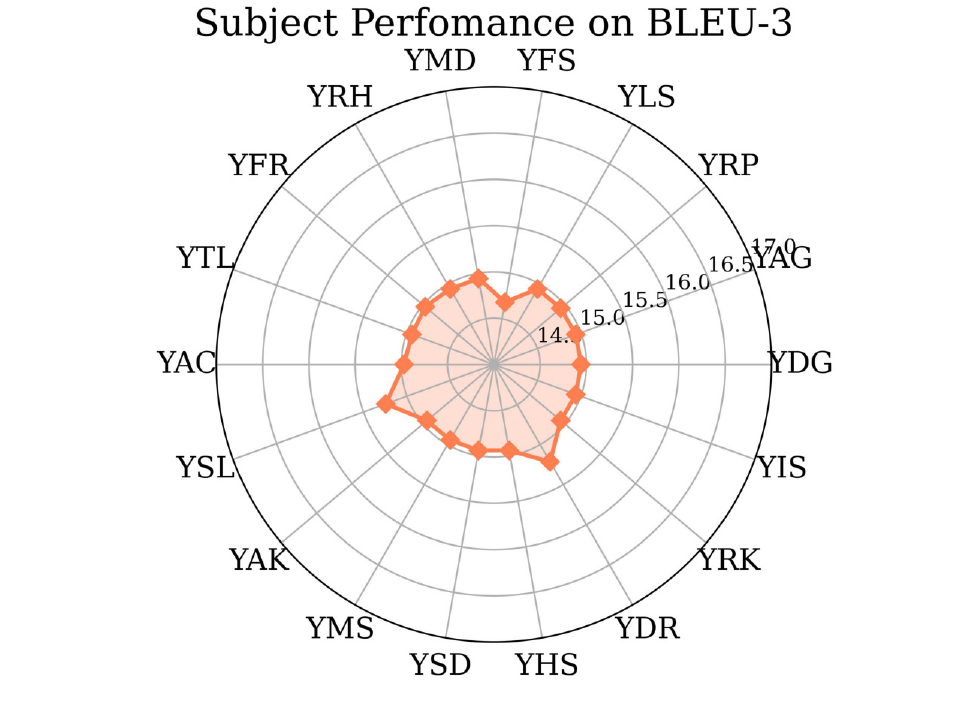}
    \end{minipage}
    \begin{minipage}[]{0.23\linewidth}
    \includegraphics[width=4cm]{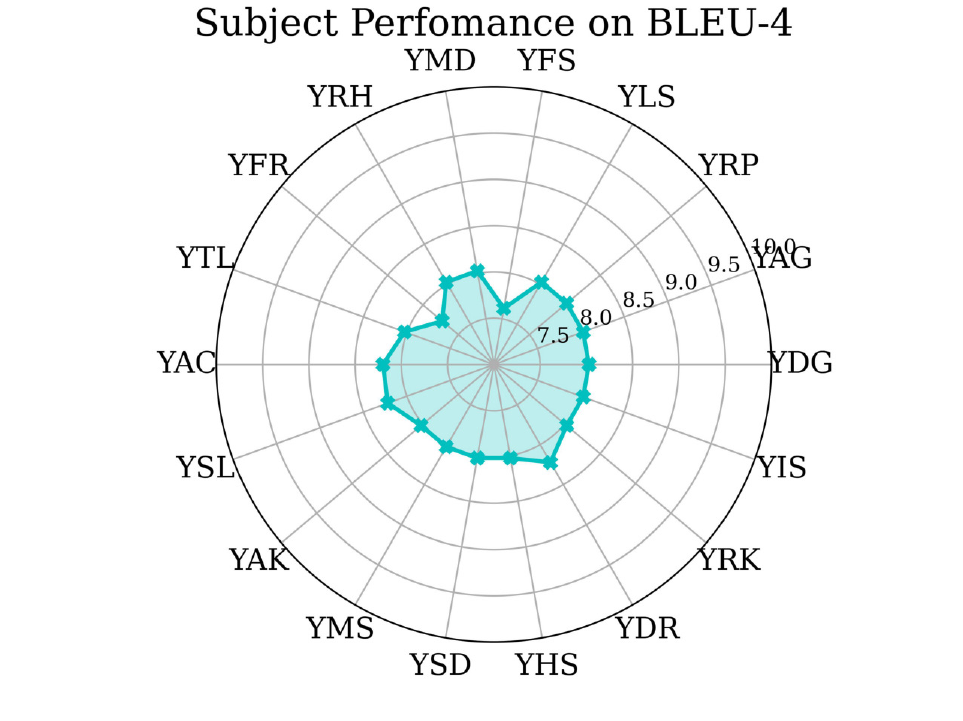}
    \end{minipage}
    \begin{minipage}[]{0.245\linewidth}
    \includegraphics[width=4cm]{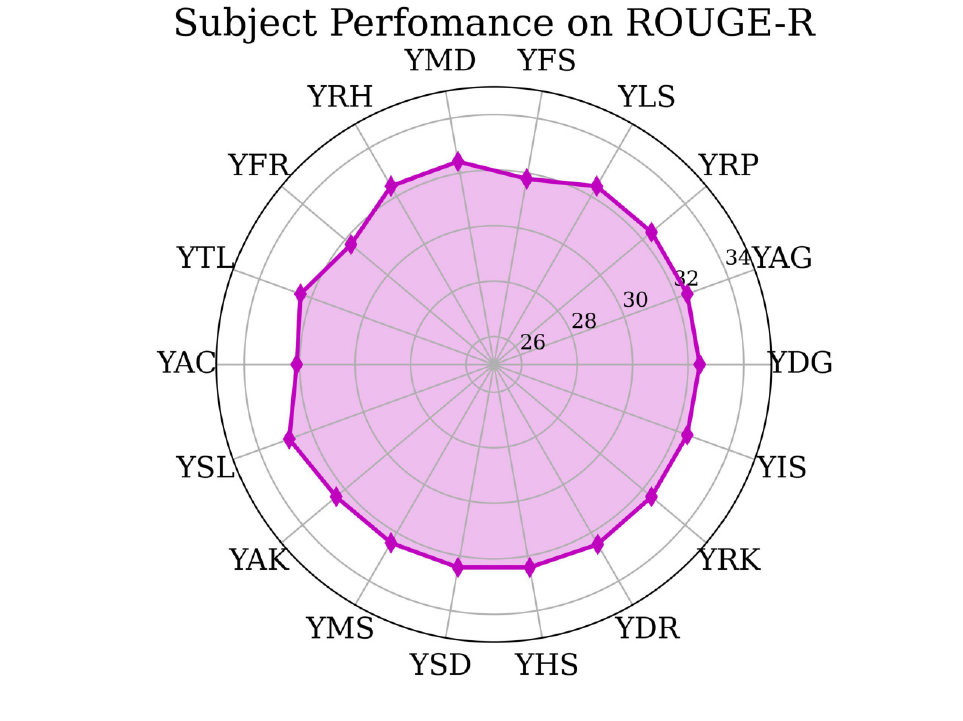}
    \end{minipage}
    \begin{minipage}[]{0.25\linewidth}
    \includegraphics[width=4cm]{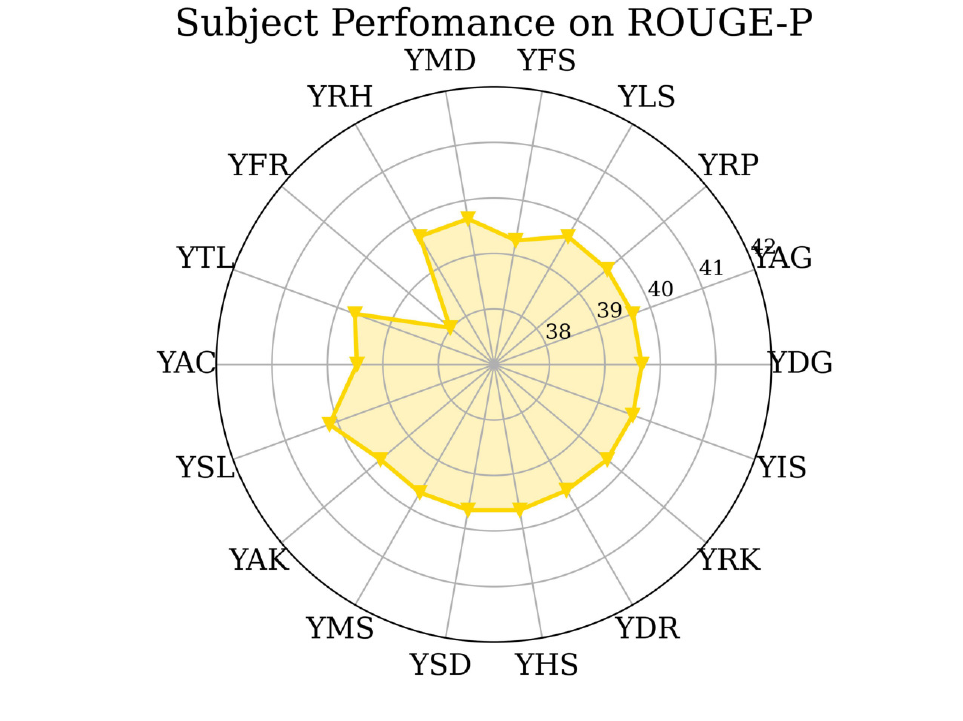}
    \end{minipage}
    \begin{minipage}[]{0.25\linewidth}
    \includegraphics[width=4cm]{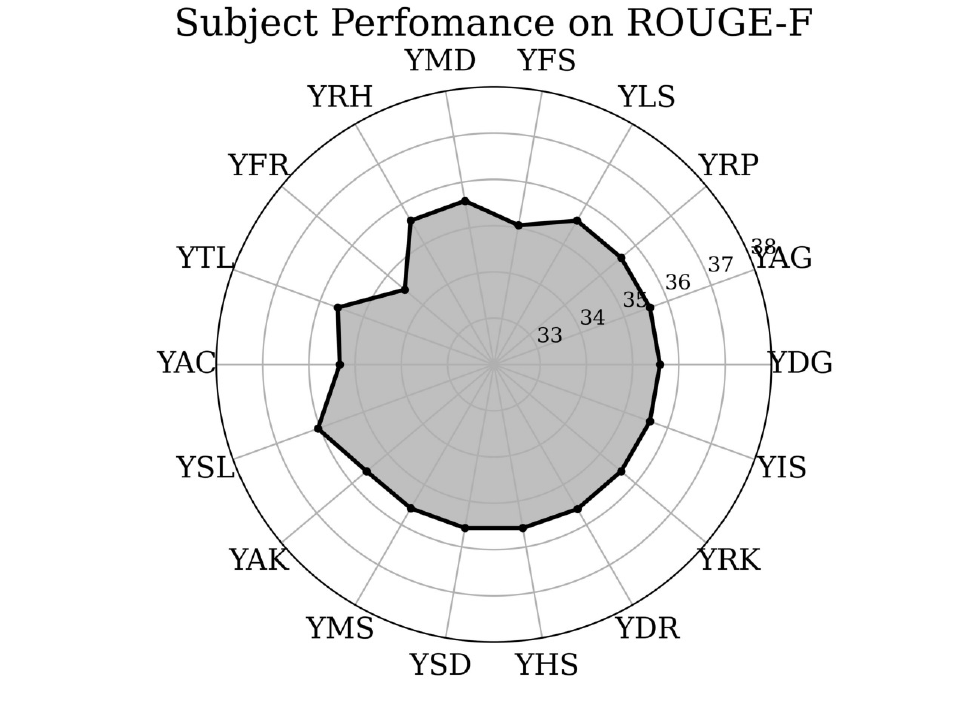}
    \end{minipage}
    \begin{minipage}[]{0.25\linewidth}
    \end{minipage}
\quad
\quad
\quad

\begin{minipage}[]{\linewidth}
\footnotesize
\resizebox{\columnwidth}{!}{
\begin{tabular}{p{1.5cm}p{0.5cm}p{0.5cm}p{0.5cm}p{0.5cm}p{0.5cm}p{0.5cm}p{0.5cm}p{0.5cm}p{0.5cm}p{0.5cm}p{0.5cm}p{0.5cm}p{0.5cm}p{0.5cm}p{0.5cm}p{0.5cm}p{0.5cm}p{0.5cm}}
\toprule
{Subject} & YDG & YAG & YRP & YLS & YFS & YMD & YRH & YFR & YTL & YAC & YSL & YAK & YMS & YSD & YHS & YDR & YRK & YIS \\ 
\midrule
{BLEU-1}  & 43.22 & 43.22 & 43.22 & 43.22 & 42.71 & 43.22 & 43.22 & 42.88 & 43.22 & 43.06 & 43.27 & 43.22 & 43.22 & 43.22 & 43.22 & 43.60 & 43.22 & 43.22 \\
{BLEU-2}   & 25.82 & 25.82 & 25.82 & 25.82 & 25.71 & 25.82 & 25.82 & 25.15 & 25.82 & 25.27 & 25.99 & 25.82 & 25.82 & 25.82 & 25.82 & 25.99 & 25.82 & 25.82 \\
{BLEU-3}& 14.70 & 14.70 & 14.70 & 14.70 & 14.46 & 14.70 & 14.70 & 14.48 & 14.70 & 14.36 & 14.81 & 14.70 & 14.70 & 14.70 & 14.70 & 14.97 & 14.70 & 14.70 \\
{BLEU-4} & 8.12 & 8.12 & 8.12 & 8.12 & 7.76 & 8.12 & 8.12 & 7.79 & 8.12 & 7.74 & 8.24 & 8.12 & 8.12 & 8.12 & 8.12 & 8.32 & 8.12 & 8.12 \\
{ROUGE-R} & 32.42 & 32.42 & 32.42 & 32.42 & 31.94 & 32.42 & 32.42 & 31.33 & 32.42 & 31.93 & 32.57 & 32.42 & 32.42 & 32.42 & 32.42 & 32.48 & 32.42 & 32.42 \\
{ROUGE-P}  & 39.71 & 39.71 & 39.71 & 39.71 & 39.40 & 39.71 & 39.71 & 38.11 & 39.71 & 39.25 & 39.84 & 39.71 & 39.71 & 39.71 & 39.71 & 39.66 & 39.71 & 39.71 \\
{ROUGE-F} & 35.65 & 35.65 & 35.65 & 35.65 & 35.23 & 35.65 & 35.65 & 34.34 & 35.65 & 35.17 & 35.79 & 35.65 & 35.65 & 35.65 & 35.65 & 35.66 & 35.65 & 35.65\\
\bottomrule
\end{tabular}}
\end{minipage}

\caption{Subject-wise evaluation results on a model trained with subject \textbf{YSL}, where the radar chart suggests the performance variance on different subjects on each metric. \label{fig:subject_wise_YSL}}
\end{table*}

\begin{table*}
    \begin{minipage}[]{0.23\linewidth}
    \includegraphics[width=4cm]{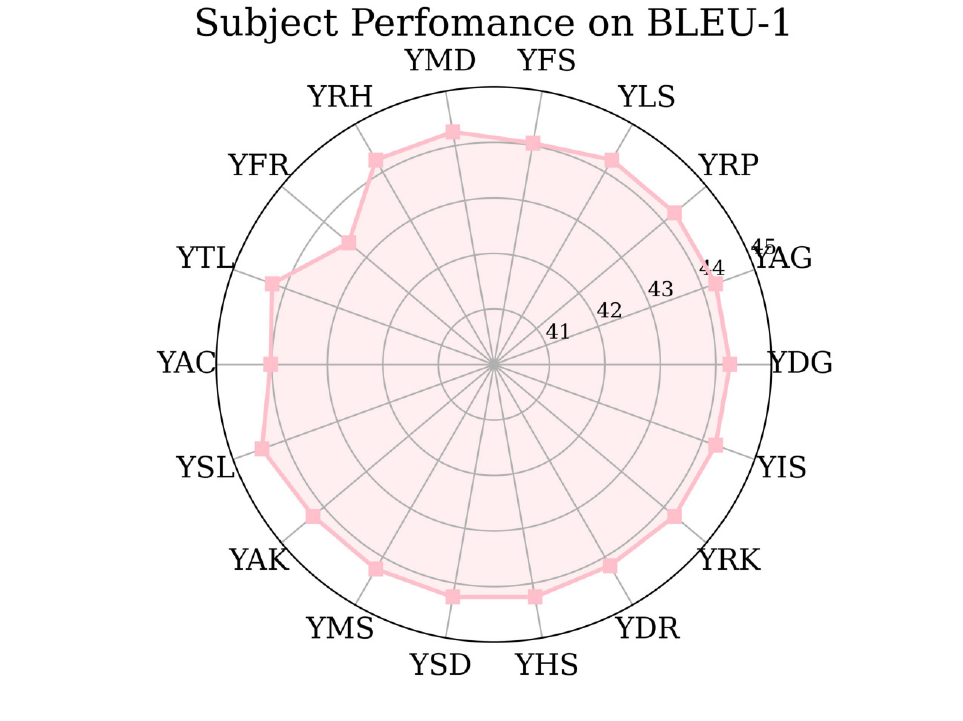}
    \end{minipage}
    \begin{minipage}[]{0.23\linewidth}
    \includegraphics[width=4cm]{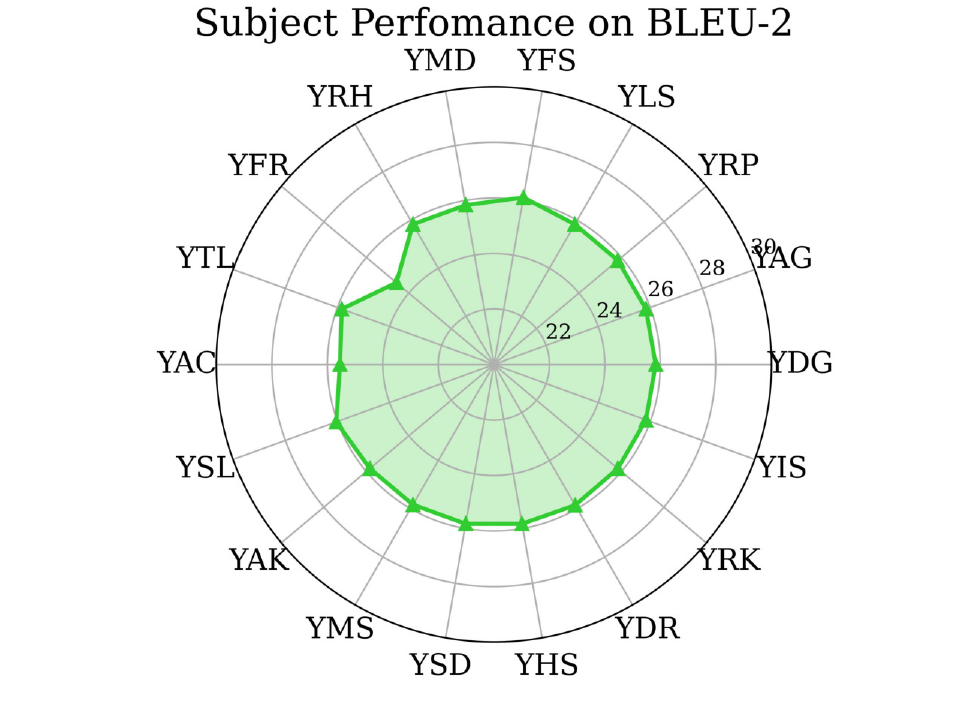}
    \end{minipage}
    \begin{minipage}[]{0.23\linewidth}
    \includegraphics[width=4cm]{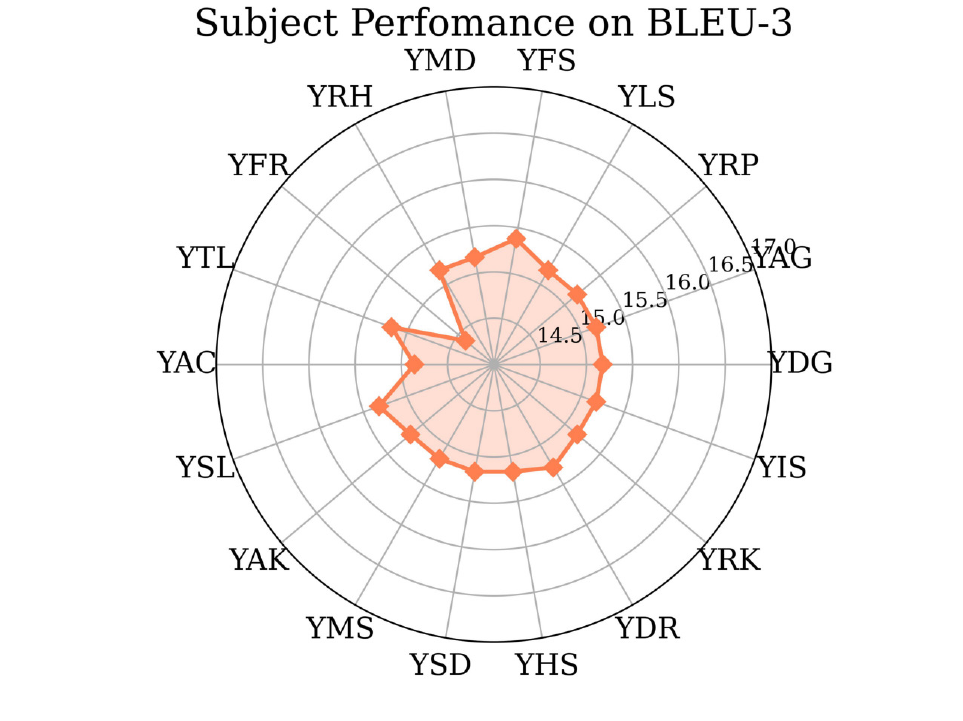}
    \end{minipage}
    \begin{minipage}[]{0.23\linewidth}
    \includegraphics[width=4cm]{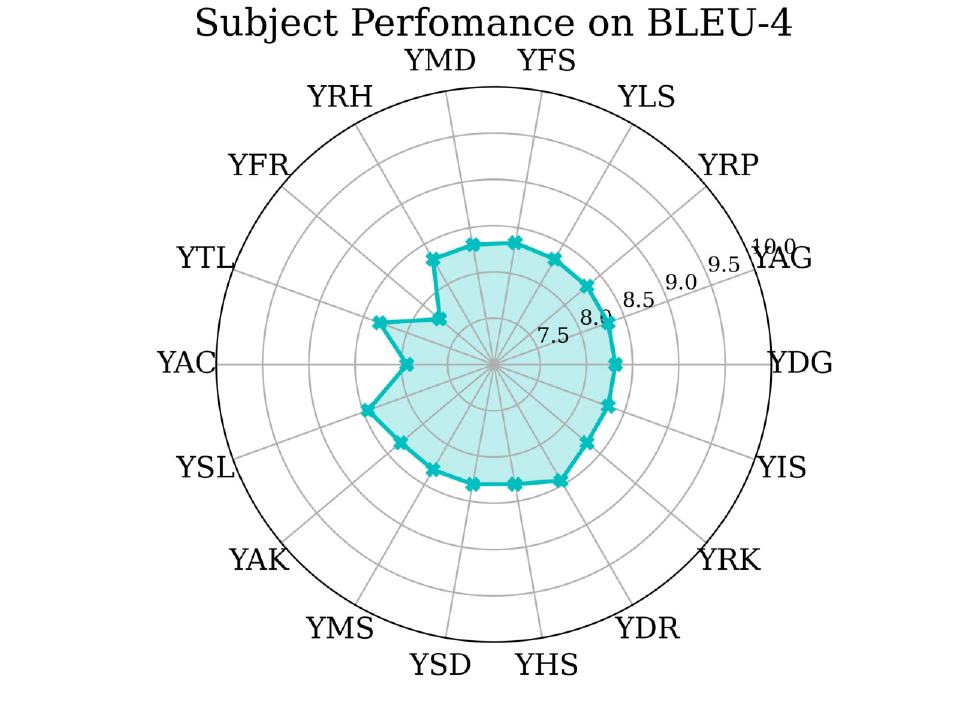}
    \end{minipage}
    \begin{minipage}[]{0.245\linewidth}
    \includegraphics[width=4cm]{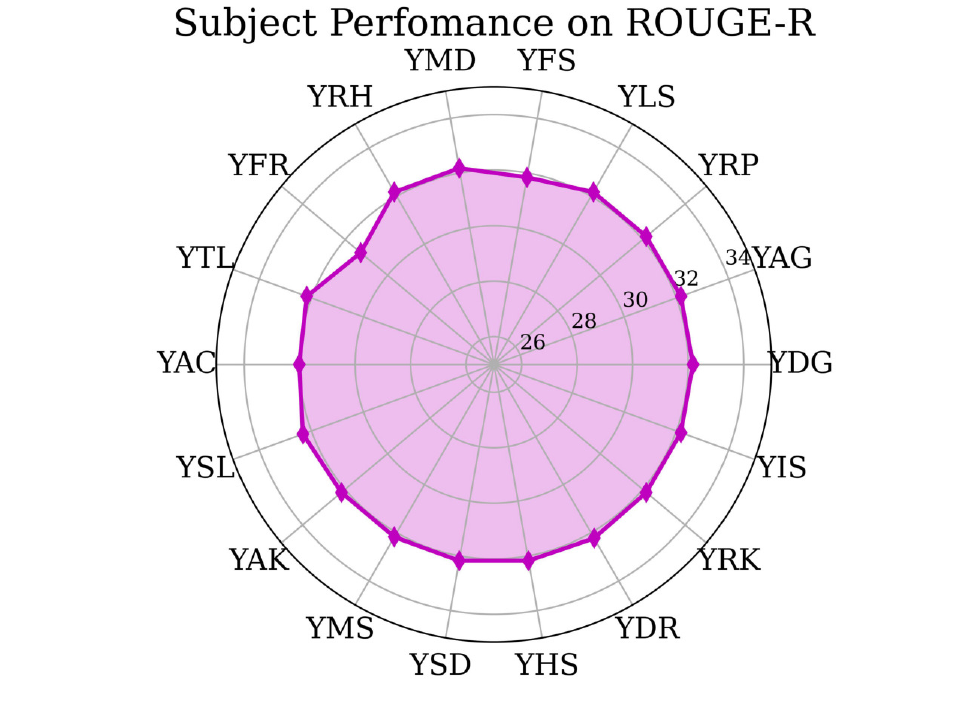}
    \end{minipage}
    \begin{minipage}[]{0.25\linewidth}
    \includegraphics[width=4cm]{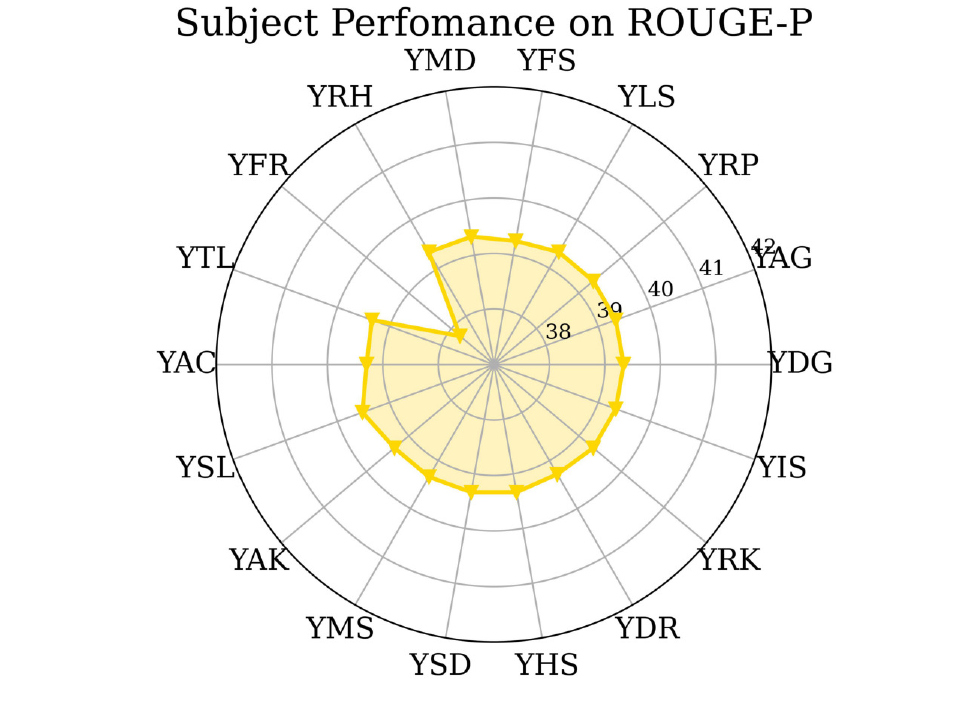}
    \end{minipage}
    \begin{minipage}[]{0.25\linewidth}
    \includegraphics[width=4cm]{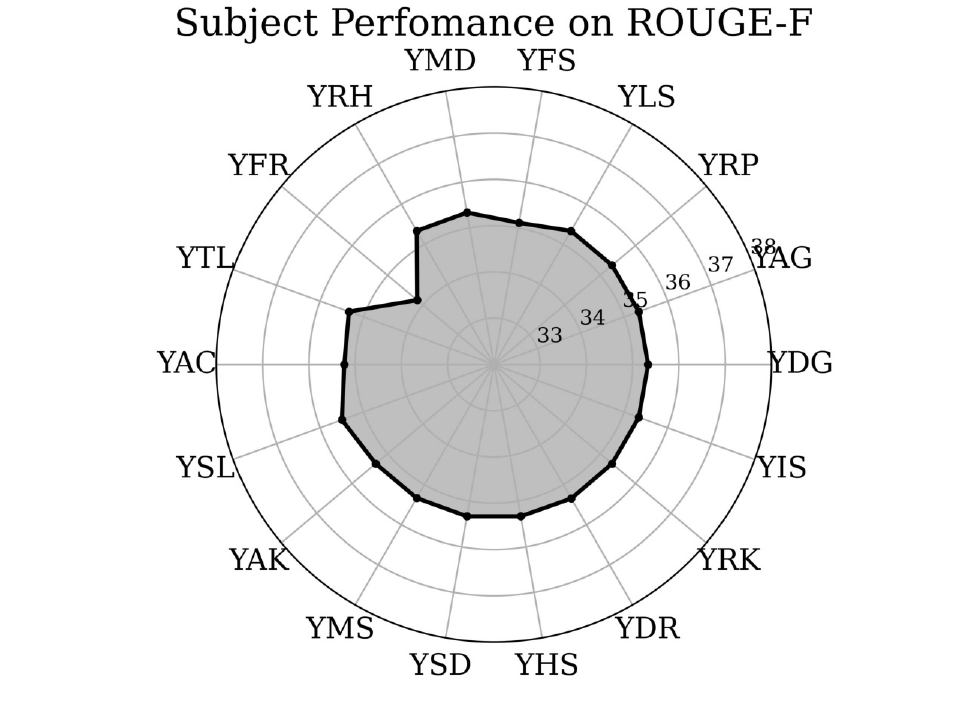}
    \end{minipage}
    \begin{minipage}[]{0.25\linewidth}
    \end{minipage}
\quad
\quad
\quad

\begin{minipage}[]{\linewidth}
\footnotesize
\resizebox{\columnwidth}{!}{
\begin{tabular}{p{1.5cm}p{0.5cm}p{0.5cm}p{0.5cm}p{0.5cm}p{0.5cm}p{0.5cm}p{0.5cm}p{0.5cm}p{0.5cm}p{0.5cm}p{0.5cm}p{0.5cm}p{0.5cm}p{0.5cm}p{0.5cm}p{0.5cm}p{0.5cm}p{0.5cm}}
\toprule
{Subject} & YDG & YAG & YRP & YLS & YFS & YMD & YRH & YFR & YTL & YAC & YSL & YAK & YMS & YSD & YHS & YDR & YRK & YIS \\ 
\midrule
{BLEU-1} & 44.25 & 44.25 & 44.25 & 44.25 & 44.05 & 44.25 & 44.25 & 43.41 & 44.25 & 44.03 & 44.45 & 44.25 & 44.25 & 44.25 & 44.25 & 44.18 & 44.25 & 44.25 \\
{BLEU-2}   & 25.83 & 25.83 & 25.83 & 25.83 & 26.11 & 25.83 & 25.83 & 24.58 & 25.83 & 25.55 & 26.04 & 25.83 & 25.83 & 25.83 & 25.83 & 25.86 & 25.83 & 25.83 \\
{BLEU-3}& 15.18 & 15.18 & 15.18 & 15.18 & 15.38 & 15.18 & 15.18 & 14.40 & 15.18 & 14.86 & 15.32 & 15.18 & 15.18 & 15.18 & 15.18 & 15.28 & 15.18 & 15.18 \\
{BLEU-4} & 8.31 & 8.31 & 8.31 & 8.31 & 8.34 & 8.31 & 8.31 & 7.76 & 8.31 & 7.94 & 8.45 & 8.31 & 8.31 & 8.31 & 8.31 & 8.45 & 8.31 & 8.31 \\
{ROUGE-R} & 32.18 & 32.18 & 32.18 & 32.18 & 31.84 & 32.18 & 32.18 & 31.27 & 32.18 & 32.02 & 32.32 & 32.18 & 32.18 & 32.18 & 32.18 & 32.23 & 32.18 & 32.18 \\
{ROUGE-P}  & 39.34 & 39.34 & 39.34 & 39.34 & 39.26 & 39.34 & 39.34 & 37.79 & 39.34 & 39.29 & 39.52 & 39.34 & 39.34 & 39.34 & 39.34 & 39.28 & 39.34 & 39.34 \\
{ROUGE-F} & 35.34 & 35.34 & 35.34 & 35.34 & 35.11 & 35.34 & 35.34 & 34.16 & 35.34 & 35.24 & 35.50 & 35.34 & 35.34 & 35.34 & 35.34 & 35.35 & 35.34 & 35.34\\
\bottomrule
\end{tabular}}
\end{minipage}

\caption{Subject-wise evaluation results on a model trained with subject \textbf{YMD}, where the radar chart suggests the performance variance on different subjects on each metric. \label{fig:subject_wise_YMD}}
\end{table*}